\documentclass[lettersize,journal]{IEEEtran}
\usepackage[caption=false,font=normalsize,labelfont=sf,textfont=sf]{subfig}
\usepackage{textcomp}
\usepackage{stfloats}
\usepackage{verbatim}
\def\BibTeX{{\rm B\kern-.05em{\sc i\kern-.025em b}\kern-.08em
    T\kern-.1667em\lower.7ex\hbox{E}\kern-.125emX}}
\usepackage{balance}
\usepackage{cite}
\usepackage{amsmath,amssymb,amsfonts,amsthm}
\usepackage{algorithmic}
\usepackage[normalem]{ulem}
\usepackage{dsfont}
\usepackage{bbm}
\usepackage{xcolor}
\usepackage{mathalpha}
\usepackage{relsize}
\usepackage{graphicx}
\usepackage{pgfplots}
\usepackage{hyperref}
\usepackage{epstopdf}
\pgfplotsset{compat = newest}
\usepackage{orcidlink}
\usepackage{enumitem} 

\usepackage{array}
\usepackage{url}

\usepackage{optidef}
\usepackage{MnSymbol}

\usepackage{schemabloc}

\usepackage[straightvoltages]{circuitikz}
\usepackage{tikz}
\usetikzlibrary{arrows,shapes}

\usetikzlibrary{babel}
\graphicspath{/img}

\newtheorem{theorem}{Theorem}

\newtheorem{definition}{Definition}

\newtheorem{remark}{Remark}

\newcommand{\F}{\mathcal{F}}
\newcommand{\T}{\mathcal{T}}
\newcommand{\N}{\mathcal{N}}
\newcommand{\I}{\mathcal{I}}

\newcommand{\Z}{\mathbb{Z}}

\newcommand{\R}{\mathbb{R}}

\newcommand{\hm}[1]{\mathcal{#1}}

\newcommand{\w}{\omega}
\newcommand{\dbar}[1]{\dot{\overline{#1}}}
\newcommand{\dc}{\mathrm{dc}}
\newcommand{\abc}{\mathrm{abc}}
\newcommand{\dq}{\mathrm{dq}}

\DeclareMathOperator{\sat}{sat}

\makeatletter
\renewcommand*\env@matrix[1][*\c@MaxMatrixCols c]{%
  \hskip -\arraycolsep
  \let\@ifnextchar\new@ifnextchar
  \array{#1}}
\makeatother

\begin{document}

\title{Harmonic control of three-phase AC/DC converter}

\author{Maxime~Grosso~\orcidlink{0009-0007-6721-781X}, Pierre~Riedinger~\orcidlink{0000-0002-8221-4736}, Jamal~Daafouz~\orcidlink{0000-0001-8313-8790}, 
Serge~Pierfrederici~\orcidlink{0000-0003-3682-6317}, Hicham~Janati~Idrissi~\orcidlink{0009-0009-4332-0657} and~Blaise~Lapotre
\thanks{Manuscript submitted July 13$^{th}$.  (\emph{Corresponding author Maxime Grosso})}
\thanks{Maxime Grosso is with Research Center for Automatic Control of Nancy (Université de Lorraine, CNRS, CRAN, F-54000 Nancy), France and also with Safran Electronics \& Defense, Research \& Technologies department, Massy, France (maxime.grosso@univ-lorraine.fr).}
\thanks{Pierre~Riedinger and Jamal~Daafouz are with Research Center for Automatic Control of Nancy (Université de Lorraine, CNRS, CRAN, F-54000 Nancy, France) (pierre.riedinger@univ-lorraine.fr, jamal.daafouz@univ-lorraine.fr)}
\thanks{Serge~Pierfrederici is with Laboratory of Energies Theoretical and Applied Mechanics (Université de Lorraine, CNRS, LEMTA, F-54000 Nancy, France) (serge.pierfederici@univ-lorraine.fr)}
\thanks{Hicham~Janati~Idrissi and Blaise~Lapotre are with Safran Electronics \& Defense, Research \& Technologies department, Massy, France (hicham.janati-idrissi@safrangroup.com;blaise.lapotre@safrangroup.com)}
}



\markboth{Journal of \LaTeX\ Class Files,~Vol.~14, No.~8, August~2021}%
{Shell \MakeLowercase{\textit{et al.}}: A Sample Article Using IEEEtran.cls for IEEE Journals}


\maketitle

\begin{abstract}
In this paper, we propose a harmonic-model based control approach for the three-phase grid-tied AC-DC converter. We derive a nonlinear harmonic domain model and demonstrate its efficacy in designing harmonic control with global stability guarantees. Furthermore, we establish that this harmonic control design can be translated into a periodic nonlinear control scheme in the time domain, maintaining the same level of stability guarantees. Additionally, the proposed framework allows to incorporate control objectives specifically related to harmonic distortion and harmonic mitigation as well as the tracking of periodic trajectories. Illustrative simulations and experimental setups were conducted to evaluate the effectiveness of the proposed methodology in reducing Total Harmonic Distortion (THD) in real-time. The obtained results demonstrate the successful achievement of the desired control objectives, validating the efficacy of the proposed harmonic control approach.
\end{abstract}

\begin{IEEEkeywords}
AC DC conversion, Harmonic Control, Active Filtering, Dynamic Phasor, Extended Harmonic Domain.
\end{IEEEkeywords}

\IEEEpeerreviewmaketitle

\section*{Introduction}\label{sect:Intro}
%
\IEEEPARstart{I}{n}   
recent years, the demand for high power quality in electrical systems has significantly increased due to the widespread use of interconnected electronic devices in microgrids. As a result, addressing the issue of harmonic distortion has become a critical concern \cite{mollerstedtOutControlBecause2000}. Harmonic distortion can lead to equipment malfunctions and efficiency losses, making its mitigation essential. Traditionally, the approach to mitigate harmonic distortion involved the use of heavy RLC filters to "eliminate" harmonic content. Another commonly employed approach is to either decrease the controllers passband or adjust the passband of physical filters. While effective to some extent, these methods often come at the expense of sacrificing system dynamics. To specifically target certain frequency components, techniques like incorporating filtering elements such as notch filters into the control loop have been employed. Additionally, proportional-resonant controllers with resonant components at desired frequencies have been widely used \cite{liserreMultipleHarmonicsControl2006}. In addition to these techniques, classical methods for harmonic control and mitigation also include the utilization of proportional (multi) resonant controllers \cite{teodorescuProportionalresonantControllersFilters2006} and generalized integrators \cite{yuanStationaryframeGeneralizedIntegrators2002}. It is worth noting that the aforementioned approaches have been extensively studied and applied in practice. However, each approach has its limitations and trade-offs.\\

In order to tackle the challenges posed by harmonic distortion, and go beyond the small signals assumptions, various techniques for harmonic control have been developed \cite{almerDynamicPhasorModel2015},\cite{chandrasekarAdaptiveHarmonicSteadyState2006},\cite{kamaldarAdaptiveHarmonicControl2020}.
Harmonic modeling, based on a sliding Fourier decomposition, has demonstrated its effectiveness in transforming periodic systems into time-invariant systems of infinite dimension. This approach simplifies the design of control systems, as any control strategy developed for time-invariant systems can be applied a priori, provided the infinite dimension is taken into account. In the literature, reference is often made to DHD (Dynamic Harmonic Domain) \cite{ricoDynamicHarmonicEvolution2003},\cite{karamiSinglePhaseModelingApproach2017}, DP modeling (Dynamic Phasor Modeling) techniques \cite{kotianDynamicPhasorModelling2016},\cite{almerDynamicPhasorModel2015}, Generalized State-Space Average (GSSA) techniques \cite{javaidArbitraryOrderGeneralized2015},\cite{sandersGeneralizedAveragingMethod1991},\cite{qinGeneralizedAverageModeling2012} and Harmonic State Space \cite{kwonHarmonicInteractionAnalysis2015},\cite{kwonFrequencyDomainModelingSimulation2017},\cite{ormrodHarmonicStateSpace2013}. 
In this work, we expand upon the foundations laid in the recent paper \cite{blinNecessarySufficientConditions2022}, which introduces a mathematical framework for harmonic modeling and control. This framework offers a rigorous approach that enables the design of harmonic control strategies with guaranteed stabilization in the time domain. It is important to note that in this context, the terms "Phasors" and "Harmonics" are almost interchangeable. The classical Fourier decomposition is employed to analyze periodic trajectories and generate scalar time-invariant quantities known as "Harmonics." On the other hand, "Phasors" are the time-varying counterparts of Harmonics, allowing for the consideration of transient behavior towards the periodic trajectories. \\

Our objective is to investigate the application of harmonic modeling and control design in the context of AC-DC converters. To the best of the authors' knowledge, limited emphasis has been placed in the existing literature on the mathematical conditions within the harmonic framework that ensure a stabilizing control design in the time domain. Since the harmonic model is of infinite dimension, truncation of the harmonic equations becomes necessary. However, it is crucial to exercise caution during the truncation process to avoid an inaccurate truncated representation. For instance, control methods that rely on a finite-dimensional truncated representation of the infinite-dimensional model, where all phasors above a certain order are suppressed \cite{javaidArbitraryOrderGeneralized2015, gaviriaRobustControllerFullbridge2005}, or where only specific phasors relevant to the given situation are considered \cite{ghitaHarmonicStateSpace2017}, do not guarantee the existence of an equivalent stabilizing control in the time domain. Even if the obtained harmonic control stabilizes the simplified truncated system, it does not necessarily ensure stability in the original time domain. In essence, the truncation method employed should exhibit consistency, which implies that the solutions of the truncated equations must converge to the solution of the original infinite-dimensional equation system as the order of truncation increases. This consistency ensures that the truncated representation accurately approximates the behavior of the original system. For instance, as demonstrated in \cite{riedingerSolvingInfiniteDimensionalHarmonic2022}, truncating a stable linear harmonic system at any order $k$ can result in an unstable truncated system, regardless of the chosen order. Consequently, performing a stability analysis by solving a Lyapunov equation on this truncated system would yield conclusions that are not applicable to the original infinite-dimensional system. To tackle this challenge, it is essential to develop a consistent approach for solving the Lyapunov equation in the infinite-dimensional setting, rather than relying on a reduced-order model of the system. This ensures that the stability analysis remains valid and applicable to the actual infinite-dimensional system. Expanding upon the results of \cite{riedingerSolvingInfiniteDimensionalHarmonic2022,riedingerHarmonicPolePlacement2022}, we have successfully adapted and designed nonlinear periodic forwarding control laws  specifically tailored for the AC/DC converter. Through these advancements, we have introduced integral action into the control framework, enabling precise targeting and mitigation of the harmonic content present in the phase current. This integration of integral action enhances the control strategy's ability to effectively manage and minimize the impact of harmonics, resulting in improved overall system performance. 

The paper is organized  as follows. In Section~\ref{sect:HmcTheory}, we present a comprehensive overview of the fundamental principles and notations used in harmonic modeling. Section~\ref{sect:HmcModConv} serves as a review of the fundamental equations of the AC/DC converter, presenting them in both the $abc$ and $dq$ reference frames, and provides the harmonic time invariant bilinear models for the AC/DC converter. 
Section~\ref{sect:HmcCtrlForBilSys} is devoted to the design of a harmonic stabilizing control law with global stabilization properties. Incorporating integral actions allows to enhance the control strategy by specifically targeting and mitigating the impact of individual harmonics. Leveraging an equivalence result established within the harmonic framework, we explicitly deduce the time domain counterparts of the proposed control strategy. Section~\ref{sect:CtrlObj} describes the control objectives for the AC/DC grid-tie controlled converter, with a particular emphasis on specific harmonic rejection. In Section~\ref{sect:TuningCtrl}, we customize and refine the harmonic control law to align with these control objectives. We also delve into the stability analysis of the proposed control approach under control saturation. 
Section~\ref{sect:PractImpl} describes the experimental setup used to evaluate the control law's performance, including details of the discrete-time implementation scheme. The experimental results, including measurements of Total Harmonic Distortion (THD), are detailed in Section~\ref{sect:ExpAss} and compared with those obtained using a conventional outer/inner PI loop scheme.

{\bf Notations: } The transpose of a matrix $A$ is denoted $A'$ and $A^*$ denotes the complex conjugate transpose $A^*=\bar A'$. The $n$-dimensional identity matrix is denoted $Id_n$. The infinite identity matrix is denoted $\mathcal{I}$. $C^a$ denotes the space of absolutely continuous function,
$L^{p}([a\ b],\mathbb{C}^n)$ (resp. $\ell^p(\mathbb{C}^n)$) denotes the Lebesgues spaces of $p-$integrable functions on $[a, b]$ with values in $\mathbb{C}^n$ (resp. $p-$summable sequences of $\mathbb{C}^n$) for $1\leq p\leq\infty$. $L_{loc}^{p}$ is the set of locally $p-$integrable functions. The notation $f(t)=g(t)\ a.e.$ means almost everywhere in $t$ or for almost every $t$. 
To simplify the notations, $L^p([a,b])$ or $L^p$ will be often used instead of $L^p([a,b],\mathbb{C}^n)$. 
\section{Harmonic modeling} \label{sect:HmcTheory}
In this paper, we consistently adopt the following definitions, which serve as a foundation for our discussions.\subsection{Basics definition, notations and results}
Let us consider a complex-valued function of time, denoted as $x \in L^{2}_{\text{loc}}(\mathbb{R},\mathbb{C})$. Its sliding Fourier decomposition over a window of length $T$ is defined by the time-varying infinite sequence $X = \mathcal{F}(x) \in C^a(\mathbb{R},\ell^2(\mathbb{C}))$ (see \cite{blinNecessarySufficientConditions2022}) whose components satisfy:
$$X_{k}(t)=\frac{1}{T}\int_{t-T}^t x(\tau)e^{-\textsf{j}\omega k \tau}d\tau$$ for $k\in \mathbb{Z}$, with $\omega=\frac{2\pi}{T}$.
If $x=(x_1,\cdots,x_n)\in L^{2}_{loc}(\mathbb{R},\mathbb{C}^n)$ is a complex valued vector function, then
$$X=\mathcal{F}(x)=(\mathcal{F}(x_1), \cdots,\mathcal{F}(x_n)).$$
\begin{definition} \label{def:phasor} 
$X_k$ is called the $k-$th phasor (harmonic) of $X$. 
\end{definition}
\begin{remark}
When $x$ is a real signal, $X_k=\operatorname{conj}( X_{-k})$,  $\forall k$.
\end{remark}

Additionally, we consider the Toeplitz transformation of a scalar function $x \in L^2_{\text{loc}}$, which is defined as an infinite-dimensional Toeplitz matrix given by:
\begin{align*}
    \mathcal{T}(x)=
    \left[
    \begin{array}{ccccc}
        \ddots & & \vdots & &\udots \\ & x_{0} & x_{-1} & x_{-2} & \\
        \cdots & x_{1} & x_{0} & x_{-1} & \cdots \\
        & x_{2} & x_{1} & x_{0} & \\
        \udots & & \vdots & & \ddots\end{array}\right],\end{align*}
where the terms $x_i$, $i\in\mathbb{Z}$ refers to the components of $X=\mathcal{F}(x)$.
For a matrix function $A=(a_{ij})_{i=1,\cdots, m,\ j=1,\cdots, n}$ ($\in L^2_{loc}(\mathbb{R},\mathbb{R}^{m\times n}$), this extends to
\begin{equation} \label{ToepFormMat}
\mathcal{A} =\mathcal{T}(A)= \begin{bmatrix}
\mathcal{A}_{11} & \ldots  & \mathcal{A}_{1n} \\ 
\vdots&&\vdots \\ 
\mathcal{A}_{m1} &\ldots & \mathcal{A}_{mn}
\end{bmatrix} 
\end{equation}
with $\mathcal{A}_{ij}=\mathcal{T}(a_{ij})$, $i=1,\cdots, m,\ j=1,\cdots, n$. 

\begin{definition}\label{H} We say that $X$ belongs to $H$ if $X$ is an absolutely continuous function (i.e $X\in C^a(\mathbb{R},\ell^2(\mathbb{C}^n))$ and fulfils for any $k$ the following condition: \begin{equation*}\dot X_k(t)=\dot X_0(t)e^{- \textsf{j}\omega k t} \ a.e.\end{equation*}
\end{definition}

Similar to the Riesz-Fisher theorem, which establishes a one-to-one correspondence between the spaces $L^2$ and $\ell^2$, the following theorem establishes a one-to-one correspondence between the spaces $L_{\text{loc}}^2$ and $H$.
\begin{theorem}[\cite{blinNecessarySufficientConditions2022}]\label{coincidence}   For a given $X\in L_{loc}^{\infty}(\mathbb{R},\ell^2(\mathbb{C}^n))$, there exists a representative $x\in L^2_{loc}(\mathbb{R},\mathbb{C}^n)$ of $X$, i.e. $X=\mathcal{F}(x)$, if and only if $X \in H$.
\end{theorem}

For harmonic control design, Theorem~\ref{coincidence} has a fundamental consequence: the design of a control $U$ in the harmonic domain must belong to the space $H$, otherwise its time domain counterpart $u$ does not exist. For example, if one attempt to design a state feedback $U=-\mathcal{K}X$ with a constant gain operator $\mathcal{K}$, it is proven that $\mathcal{K}$ must be a Toeplitz block operator \cite{blinNecessarySufficientConditions2022}. Hence, $u$ is determined by $u=-Kx$ where $K=\sum_{k\in\mathbb{Z}} K_ke^{{\rm j}\omega kt}$. 
Furthermore, for the purpose of ensuring bounded control $u$, it is necessary for $\mathcal{K}$ to be a bounded operator on $\ell^2$. It is worth noting that $\hm A=\mathcal{T}(A)$ is a bounded operator on $\ell^2$ if and only if $A$ is a matrix function that belongs to $L^\infty([0, T])$. Fortunately, advancements within the harmonic framework have led to the development of methods that can design controls possessing all the necessary properties. These methods include harmonic Lyapunov, Sylvester, and Riccati techniques \cite{riedingerSolvingInfiniteDimensionalHarmonic2022},\cite{ riedingerHarmonicPolePlacement2022}, as well as harmonic LMI (Linear Matrix Inequality) tools \cite{vernereySolvingInfinitedimensionalToeplitz2023}. One crucial aspect of these methods is their ability to account for the infinite dimensionality of harmonic equations while ensuring the practical determination of solutions with arbitrarily small errors. This is essential to guarantee stabilization properties. 
We recall, in the next subsection, harmonic modelling of a general differential equation.
\subsection{Harmonic modeling of dynamical systems}
By invoking Theorem~\ref{coincidence}, any system with solutions in the Caratheodory sense can be transformed through a sliding Fourier decomposition into an infinite-dimensional system. This transformation establishes a direct mapping between the trajectories of the original system and the trajectories within the infinite-dimensional space, given that the trajectories in the infinite-dimensional space belong to the subspace $H$. Moreover, when considering a system that is periodic with period $T$, the resulting infinite-dimensional system becomes time-invariant. 
For instance, under a weak assumption (see \cite{blinNecessarySufficientConditions2022}), any n-dimensional differential system: $$\dot x=f(t,x)$$ having Caratheodory solution has an harmonic representation given by$$\dot X=\mathcal{F}(f(t,x))-\mathcal{N}X.$$
where \begin{equation}\mathcal{N}=\mathrm{Id}_n\otimes \mathrm{diag}( \textsf{j}\omega k,\ k\in \mathbb{Z})\label{q}\end{equation}
Moreover $x$ can be recovered from $X$ thought the exact formula\cite{blinNecessarySufficientConditions2022}:
\begin{equation}\label{inv_four_reconstruc} 
x(t)=\mathcal{F}^{-1}(X)(t)=\sum_{k=-\infty}^{+\infty} X_k(t)e^{ \textsf{j}\omega k t}+\frac{T}{2}\dot X_0(t)
\end{equation}
For polynomial systems, the term $\mathcal{F}(f(t,x))$ can be easily and explicitly determined using the following rule:
\begin{enumerate}
    \item For a $A$ matrix function ($\in L_{loc}^\infty$) and a vector $x\in L_{loc}^2$:
\begin{equation}
  \F\left(Ax\right)=\mathcal{T}(A)\mathcal{F}(x)=\mathcal{A}X
\end{equation}
where $X=\mathcal{F}(x)$ and where $\mathcal{A}=\mathcal{T}\left(A\right)$.
\item For two matrix functions $A$ and $B$ ($\in L_{loc}^\infty$):
\begin{equation}
  \mathcal{T}\left(AB\right)=\mathcal{T}\left(A\right)\mathcal{T}\left(B\right)=\mathcal{AB}
\end{equation}
\end{enumerate}
It is important to highlight that for a $T$-periodic vector $x$ or matrix function $A$, both $X=\mathcal{F}(x)$ and $\mathcal{A}$ are constant. Specifically, if $A$ is a constant matrix, we obtain the following relationship:
\begin{equation}
  \mathcal{A}=\mathcal{T}\left(A\right)= A \otimes\mathcal{I}
\end{equation}
with $\mathcal I$ the infinite identity matrix and $\otimes$ the Kroenecker product. {When considering a lowercase vector-valued function $x \in L^{2}_{\text{loc}}(\mathbb{R},\mathbb{C}^n)$, we adopt the following convention: $X = \mathcal{F}(x)$ represents the Fourier transform of $x$, while $\hm X = \T (x)$ represents the Toeplitz form of $x$. By abuse of notation, we also use $\T(X)$ in place of $\T(x)$ to denote the function that constructs a Toeplitz matrix using the elements of $X=\mathcal{F}(x)$.}

\section{Harmonic modeling of the AC/DC grid tied converter}\label{sect:HmcModConv}
Our objective is to investigate the application of harmonic modeling and control design in the context of AC-DC converters. We start by recalling the dynamical equations for a classical topology of the AC-DC converter in the time domain, considering both the $abc$ and $dq$ reference frames.
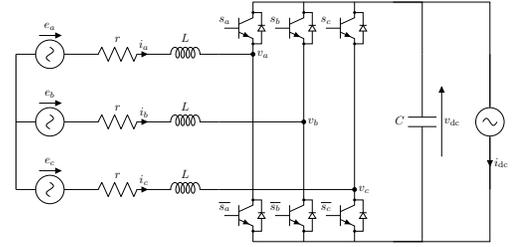
\begin{figure}[!t]
\centering \scalebox{0.45}{
\begin{circuitikz}
\draw (0,4) to[sV=$e_c$] ++(2,0) to[R=$r$,i=$i_c$] ++(2,0) to[L=$L$] ++(2,0);
\draw (0,6) to[sV=$e_b$,name=eb] ++(2,0) to[R=$r$,i=$i_b$] ++(2,0) to[L=$L$] ++(2,0);
\draw (0,8) to[sV=$e_a$] ++(2,0) to[R=$r$,i=$i_a$] ++(2,0) to[L=$L$] ++(2,0);
\draw (0,4) -- (0,8);
\draw (6,4) -- (10,4) node[circ]{} ++(0,0) node[right] {$v_c$};
\draw (6,6) -- (8.5,6) node[circ]{} ++(0,0) node[right] {$v_b$};;
\draw (6,8) -- (7,8) node[circ]{} ++(0,0) node[right] {$v_a$};;
\draw (7,8) node [npn, bodydiode, anchor=E](ta1){} to ++(0,-4) node [npn, bodydiode, anchor=C](ta2){} ;
\draw (8.5,8) node [npn, bodydiode, anchor=E](tb1){} to ++(0,-4) node [npn, bodydiode, anchor=C](tb2){} ;
\draw (10,8) node [npn, bodydiode, anchor=E](tc1){} to ++(0,-4) node [npn, bodydiode, anchor=C](tc2){} ;
\draw (ta1.C) -- (tb1.C) -- (tc1.C) to[short,name=sh1] ++(2,0) to[short] (12,8);
\draw (ta2.E) -- (tb2.E) -- (tc2.E);
\draw (tc2.E) to[short,name=sh2] ++(2,0) to [short] (12,4) to[C=$C$,name=C,v=$v_\dc$] ++(0,4) ;
\draw (14,8) to[sI=$i_\dc$,name=sI] (14,4);
\draw (sh2) -| (sI) |- (sh1);
\draw (ta1.B) node[above] {$s_a$};
\draw (tb1.B) node[above] {$s_b$};
\draw (tc1.B) node[above] {$s_c$};
\draw (ta2.B) node[above] {$\overline{s_a}$};
\draw (tb2.B) node[above] {$\overline{s_b}$};
\draw (tc2.B) node[above] {$\overline{s_c}$};
\end{circuitikz}}
\caption{schematic of a grid tied AC/DC converter with load represented as current source}
\label{fig:1_tikz_ACD}
\end{figure}
The AC-DC converter's topology is illustrated in Figure~\ref{fig:1_tikz_ACD}. For simplicity, we consider a single $L$ filter with a line resistance $r$. The load is represented as a current source denoted by $i_\dc$. We define $s_i$ as the switching variable for arm $i$, taking values of $0$ or $1$.  
Hence, considering the associated duty cycle $d_i$, defined in $[0,1]$ as the average value of $s_i$ over a switching period, and a balanced operational mode ($e_a+e_b+e_c=0$), the average model under the $abc$ reference frame is given by~\cite{zhouModelingControlThreePhase2018}:

\begin{equation} \label{model:converter_detailed_abc_time}
  \begin{cases}
    L \dfrac{\mathrm d}{\mathrm{dt}} i_\abc = -r  i_\abc  - C_{33} d_\abc  v_\dc -  e_\abc  \\
    ~~\\
    C \dfrac{\mathrm d}{\mathrm{dt}} v_\dc =  d_\abc'i_\abc  - i_\dc \end{cases}
\end{equation}
where the generic notation $x_\abc$ stands for $x_\abc=[x_a,x_b,x_c]^T$ and where the Laplacian matrix $C_{33}$ is given by: 
$$\begin{bmatrix}
\frac{2}{3} & -\frac{1}{3}& -\frac{1}{3}\\ -\frac{1}{3} & \frac{2}{3} & -\frac{1}{3} \\  -\frac{1}{3} & -\frac{1}{3} & \frac{2}{3}
\end{bmatrix}.$$ 
It is worth noting that the row sums of this matrix equal zero, ensuring the fulfillment of Kirchhoff's law ($i_a + i_b + i_c = 0$) when the system operates in a balanced mode ($e_a + e_b + e_c = 0$).
The given system can be expressed as a bilinear system in the following generic form:
\begin{equation} 
\dot x = Ax + G(x) d + B v \label{model:converter_bilmodel_abc_time}
\end{equation}
where $x=[i_\abc,v_\dc]^T$, $v=[e_\abc,i_\dc]^T$, $d=d_\abc$ the switching duty cycle control and
$$A = \begin{bmatrix}
    -\frac{r}{L} I_3 & 0 \\ 0 & 0 
\end{bmatrix},~~~
G(x) = \begin{bmatrix}
\frac{C_{33}}{L} v_\dc \\ \frac{i_\abc^T}{C}
\end{bmatrix}, ~~~B = \begin{bmatrix}
    \frac{I_3}{L} & 0\\ 0 &-\frac{1}{C}
\end{bmatrix}.$$

For the sake of comparison, let us recall the well-established $dq$ model derived using Park's transformation. In this model, we introduce $\omega$ as the pulsation of the source grid voltage: 
\begin{equation}\label{model:converter_detailed_dq_time}
\begin{cases}
L \dfrac{\mathrm d}{\mathrm{dt}}i_\dq  = - r i_\dq  - L R  \omega i_\dq- d_\dq v_\dc - e_\dq \\ 
~\\
C \dfrac{\mathrm d}{\mathrm{dt}} v_\dc = \frac{3}{2} d_\dq' i_\dq -i_\dc\end{cases}
\end{equation}
where the generic notation $x_\dq$ stands for $x_\dq=(x_\mathrm{d},x_q)$ and $R=\begin{bmatrix}0 & -1 \\ 1 & 0 \end{bmatrix}$.

To derive the harmonic model of the the AC/DC grid-tied converter, we consider the bilinear system~\eqref{model:converter_detailed_abc_time} corresponding to the $abc$ frame. Building upon the previous section, it can be shown straightforwardly that the harmonic model can be expressed as follows:

\begin{equation*}\label{model:converter_detailed_abc_harmo}
    \begin{cases}
        L\dfrac{\mathrm d}{\mathrm{dt}} I_\abc &= I_3 \otimes (-{r}\I - L \hm N) I_\abc + (C_{33}\otimes \hm V_\dc) D +  E_\abc \\ 
        ~\\
         C \dfrac{\mathrm d}{\mathrm{dt}} V_\dc &= - C \N V_\dc +  \hm I_\abc ^* D - I_\dc
    \end{cases}
\end{equation*}

For the generic equation~\eqref{model:converter_bilmodel_abc_time}, the harmonic model reads:  
\begin{equation}\label{model:converter_bilmodel_abc_harmo}
\dot X = \left(\mathcal A - \N \right) X + \mathcal{G}\left(X\right)D + \mathcal{B}V
\end{equation}
with 
$$X=(I_a, I_b , I_c , V_\dc)=\mathcal{F}(x),$$ $$V=(E_a,E_b,E_c,I_\dc)=\mathcal{F}(e_a,i_\dc),$$ and the Harmonic Domain switching control $$D = (D_a, D_b , D_c)=\mathcal{F}(d_\abc)$$ and where 
$$
\mathcal{A}=\mathcal{T}(A) = \begin{bmatrix}  -\frac{r}{L} I_3 \otimes  \I & 0 \\ 0 & 0  \end{bmatrix}, \quad 
\hm B = \begin{bmatrix} \frac{I_3\otimes \mathcal{I}}{L}& 0 \\ 0 &   -\frac{\mathcal{I}}{C}\end{bmatrix},$$
$$
\mathcal{G}(X)=\mathcal{T}(G(x))=\begin{bmatrix}
 \frac{C_{33}  \otimes \hm V_\dc}{L} \\ \frac{\hm I_\abc^*}{C}
\end{bmatrix}$$ with $\mathcal{V}_\dc=\mathcal{T}(v_\dc)$, $\mathcal{I}_\abc=\mathcal{T}(i_\abc)$.
The period $T=\dfrac{2 \pi}{\omega}$ is naturally chosen as the greatest period of the periodic input signal $e_\abc$.
An important property that holds for a $T$-periodic signal $u(\cdot)$ is the following: $$\F(u(t-\alpha T)) = S_{-2 \pi\alpha}\F(u(t))$$
where $\hm S_\alpha = \mathop{\mathrm{diag}}(e^{j k \alpha},{k \in \mathbb{Z}})$. Therefore, under balanced operation and as in \cite{karamiSinglePhaseModelingApproach2017}, we can simplify the expression of $E_{abc}$ as a single input following the relation:
 $E_{abc}=\hm S E_a$ where $\hm S= (\hm I,S_{-\frac{2 \pi}{3}},S_{\frac{2 \pi}{3}})$. \\
 
 Similarly, the harmonic model corresponding to the AC/DC grid-tied converter in the $dq$ frame can be expressed as follows:
\begin{equation}
 \begin{cases}
 \begin{aligned}
\dfrac{\mathrm d}{\mathrm{dt}} I_\dq =   \left(-\frac{r}{L} I_2  \otimes \I   + R \otimes \hm W - \N \right) I_\dq\\ - \frac{\hm  V_\dc}{L} D_\dq- \frac{1}{L}E_\dq\end{aligned} \\ 
\dfrac{\mathrm d}{\mathrm{dt}} V_\dc =-\N V_\dc+\frac{3}{2C}\hm I_\dq D_\dq -\frac{1}{C}I_\dc\end{cases}
 \end{equation} where $\hm W = \T (\omega)$ and is equals to $\hm W= \omega \otimes \mathcal{I}$ for constant $\omega$.

Some significant relationships between the harmonic modeling in the $abc$ frame and the $dq$ modeling are worth mentioning. The fundamental harmonic is characterized by ${<I_{a}>_{1}=\frac{1}{\sqrt{6}}(i_{d,0} + j i_{q,0})}$. For established setpoint, a $k$-order harmonic on $i_\dq$ implies $(k\pm 1)$-order harmonics on $I_\abc$.


 In the next section, we discuss harmonic model-based control and introduce a generic harmonic control design framework for bilinear systems.

\section{Harmonic control for bilinear systems}\label{sect:HmcCtrlForBilSys}
As emphasized in Theorem~\ref{coincidence}, the harmonic control $D$ must belong to the space $H$. Failure to meet this condition means that the resulting control does not have a counterpart $d$ in the time domain. For example, when considering a finite-dimensional truncated linear system, where all phasors of order greater than $k$ are removed, and attempting to solve a finite-dimensional Linear Quadratic Regulator (LQR) problem, the resulting state feedback control does not belong to $H$ since the matrix gain lacks a Toeplitz block structure. Moreover, even if a Toeplitz structure is enforced, it is insufficient to prove that the resulting control in the time domain is stabilizing. Once again, demonstrating the consistency of the truncation scheme becomes crucial to establish such a result. It is important to highlight that the harmonic Lyapunov-Sylvester based control method proposed in this section incorporates consistent schemes, as demonstrated in previous works \cite{riedingerSolvingInfiniteDimensionalHarmonic2022, vernereySolvingInfinitedimensionalToeplitz2023, riedingerHarmonicPolePlacement2022}. Consequently, the existence and stabilizing properties of the control in the time domain are guaranteed by the equivalence between time domain trajectories and infinite-dimensional harmonic trajectories. This section specifically focuses on the harmonic control design for bilinear systems of the form presented in~\eqref{model:converter_bilmodel_abc_time}. To this end, we consider an equilibrium $(X^e,D^e,V^e)$ for the system (\ref{model:converter_bilmodel_abc_harmo}), given by:
 \begin{equation} \label{harmeq}
0 = (\mathcal{A} - \mathcal{N})X^e+ \hm G(X^e)D^e+ \mathcal{B} V^e
\end{equation}
Let the state error $\overline X=X-X^e$, $\overline D=D-D^e$ and $\mathcal{D}^e=\mathcal{T}(D^e)$. The error dynamics reads:
 \begin{equation}\label{model:errordyn_abc_harmo}
   \dbar X = (\mathcal{A} + \mathcal{A}(\mathcal{D}^{e}) - \mathcal{N})\overline X + \mathcal{G}(X)\overline D
 \end{equation} where $$\mathcal{A}(\mathcal{D}^{e})=\begin{bmatrix} 0& 
 (\frac{C_{33}  \otimes \hm I )\hm D^e}{L} \\ \frac{\hm D^{e*}}{C} & 0
\end{bmatrix}.$$ 
Our objective is to design a stabilizing feedback law for the system given by~\eqref{model:errordyn_abc_harmo}. We approach this task in two steps.
In the first step, we propose a method to determine a stabilizing feedback law. 
In the second step, we enhance this feedback law by incorporating forwarding control methods \cite{mazencAddingIntegrationsSaturated1996,simonRobustRegulationPower2023} adapted for the harmonic domain. Indeed, incorporating harmonic integral actions allows to address harmonic disturbance rejection and trajectory tracking.

\begin{remark} Let $\overline{x}= x(t) - x^e(t)$ and $\overline{d}=d(t)-d^e(t)$. 
    The error model can be expressed in the time domain as a periodic system: \begin{equation}
        \dot{\overline{x}} = (A + A(d^e))\overline{x} +G(x) \overline{d}\label{model:errordyn_abc_time}
    \end{equation}  where $A,G$ are given by~\eqref{model:converter_bilmodel_abc_time} and $$A(d^e) = \begin{bmatrix}
      0 &\frac{C_{33}  d^e}{L} \\ \frac{ d^{e'}}{C} & 0
    \end{bmatrix}.$$ 
    The harmonic control designed to stabilize  \eqref{model:errordyn_abc_harmo} must also stabilize \eqref{model:errordyn_abc_time} using the inverse sliding Fourier decomposition defined in \eqref{inv_four_reconstruc}. 
  
\end{remark}

\subsection{Stabilizing Harmonic State feedback} 
The following result allows to design a state feedback harmonic control with stability guarantees. 
\begin{theorem}\label{stab} Assume $(\hm{A}+\hm A( \hm D^e)- \hm{N})$ is an Hurwitz operator and consider a symmetric and positive definite matrix function $Q\in L^\infty([0\ T])$ and $\hm Q=\hm T(Q)$. Then, the solution $\hm P$ of the harmonic Lyapunov equation:
    \begin{equation} 
     (\hm{A}+\hm A( \hm D^e)- \hm{N})^* \hm P + \hm P (\hm{A}+\hm A( \hm D^e)- \hm{N}) + \hm Q = 0 \label{lyapunov:hm}
\end{equation}
is a hermitian, positive definite and bounded operator on $\ell^2$.

Moreover, for any positive definite, symmetric matrix function $H_1\in L^\infty([0\ T])$, the state feedback control law given by:
\begin{equation}
    \label{control:SF_hmq} {D} = D^{e} - \mathcal{H}_1 \mathcal{G}^*(X) \mathcal{P}(X-X^{e})
\end{equation}
    where $\hm H_1=\hm T(H_1)$
    stabilizes globally and asymptotically the state $X$ of system~\eqref{model:converter_bilmodel_abc_harmo} to $X^e$.
\end{theorem} 
A complete proof of this theorem, based on the Lyapunov function $V(\overline X)=\overline X^*\hm P\overline X$, is provided in \cite{blinNecessarySufficientConditions2022}. From a practical point of view, an efficient method for solving the infinite-dimensional equation~\eqref{lyapunov:hm} with arbitrary precision is presented in \cite{riedingerSolvingInfiniteDimensionalHarmonic2022}, allowing for effective implementation of the control. Furthermore, it should be noted that the matrix $H_1$ serves as a design parameter, allowing for adjustment of the control's aggressiveness. By tuning the values of $H_1$, the control can be tailored to achieve the desired performance characteristics.
 Now, as a consequence of Theorem~\ref{coincidence}, as $D$ belongs trivially to the space $H$, it has a time domain counterpart $d$ for system~\eqref{model:converter_bilmodel_abc_time}. This time domain counterpart can be obtained by applying formula~\eqref{inv_four_reconstruc}, resulting in the following expression:
\begin{equation}\label{control:stabilizing_time}
d = d^{e} - H_1  G(x)^T P(x-x^e)  
\end{equation}
with the same guaranteed stability properties.
Notice that $H_1=\T^{-1}(\hm H_1)$ and $P=\T^{-1}(\hm P)$ are periodic gain matrices.

\subsection{Harmonic Forwarding Control}
The previous control strategy can be enhanced by incorporating integral actions to mitigate the impact of constant perturbations in the harmonic domain or, equivalently, $T-$periodic perturbations in the time domain. Through the incorporation of integral actions, we can guarantee, for example, the regulation of the mean value of the voltage $v_{dc}$, even in the presence of unknown load current $i_{dc}$ and effectively reject harmonic disturbances. To further advance our objective, we introduce additional equations to both system~\eqref{model:converter_bilmodel_abc_time} and~\eqref{model:errordyn_abc_harmo}. These equations, denoted as~\eqref{forwarding:OLC_time} and~\eqref{forwarding:OLC_hmq}, respectively, are defined as follows:
\begin{equation} \label{forwarding:OLC_time}
\dot z =O z + LC(x-x^{e})
\end{equation}
\begin{equation} \label{forwarding:OLC_hmq}
\dot Z =( \mathcal{O} - \mathcal{N})Z + \mathcal{L}\mathcal{C}(X-X^{e})
\end{equation}
Here, $Z=\hm F(z)$, the matrix $O$ is skew-symmetric and belongs to the space $L^\infty([0 \ T])$ as a matrix function, and $\hm{O}=\hm T(O)$. Similarly, $L$ is a matrix function in $L^\infty([0 \ T])$ and $\hm{L}=\hm T(L)$. The matrix $C$ is used as an output matrix to select specific components of $X$, and $\mathcal{C}=\hm T(C)$. 

Consider $\hm M$ as the solution to the harmonic Sylvester equation given by:
\begin{equation} \label{Sylvester:hm} \mathcal{(O-N)M} -\hm{M}(\mathcal{A + A(D}^{e})-\mathcal{N}) + \mathcal{LC} = 0 \end{equation}
It is important to note that $\hm M$ is a bounded operator on $\ell^2$ or, equivalently, its time-domain counterpart $M$ is a matrix function in $L^\infty([0\ T])$ {that solves the time domain periodic differential Sylvester equation: \begin{equation}
    \dot M(t) - O M(t) + M(t) ( A + A(d^e(t)) - LC(t) =0 \end{equation}}
\begin{theorem} Using $\hm P$ as provided by Theorem \ref{stab} and for any symmetric and positive definite matrix functions $H_1$ and $H_2\in L^\infty([0\ T])$ {such that $ H_2 O - O H_2 = 0$}, the state feedback control law given by:
\begin{equation}    \label{control:hmq_form}
    \begin{aligned}
        D = D^{e} - \hm H_1 \mathcal{G}^*(X) [&\mathcal{P}(X-X^{e})\\& -  \mathcal{M}^* \hm H_2(Z- \mathcal{M}(X-X^{e}))]
    \end{aligned}
    \end{equation}
where $\hm H_i=\hm T(H_i)$, $i=1,2$ stabilizes globally and asymptotically the state $X$ of system~\eqref{model:converter_bilmodel_abc_harmo} to $X^e$.
\end{theorem} 
The complete proof of the theorem can be found in \cite{blinNecessarySufficientConditions2022} and an efficient algorithm is provided in \cite{riedingerHarmonicPolePlacement2022} to practically solve the harmonic Sylvester equation up to an arbitrarily small error.
Finally, since $D$ belongs to the space $H$, it has a time-domain counterpart for system~\eqref{model:converter_bilmodel_abc_time}, which can be directly obtained as follows:
\begin{equation}    \label{control:time_form}
    \begin{aligned}
        d = d^{e} -  H_1 G^T(x) [&P(x-x^{e})\\& - M^TH_2(z- M(x-x^{e}))]
    \end{aligned}
    \end{equation}
In the upcoming section, we delve into a comprehensive explanation of the design and practical application of this generic stabilizing control law. We outline the steps and considerations involved in adapting this control law to accomplish our specific control objectives effectively.

\section{Control objectives for the ac/dc rectifier}\label{sect:CtrlObj}
We consider the three-phase AC/DC rectifier with numerical values given in Table \ref{tab:numerical_elec_value}. These values are representative of the actual parameters of our experimental setup.
We first provide an overview of the control objectives within the context of our AC/DC rectifier. Subsequently, we establish a nominal periodic set-point for the system, which corresponds to the desired operating mode under normal conditions.
\subsection{Objectives}
We consider the following objectives:
\begin{enumerate}
    \item Primary objectives: under T-periodic load or grid perturbation:
    \begin{itemize}
\item maintain the DC bus voltage mean value at a given reference $v_{dc_{ref}}$,
\item maximize power factor: $i_{q}$ mean value must be maintained to 0.
\end{itemize}
    \item Secondary objective: reject $2$-nd and $4$-th order harmonic in $i_\abc$,  due to $3$-rd order harmonic perturbation in $i_\dc$.
\end{enumerate}
Certainly, the secondary objective serves as an illustrative example to demonstrate the capabilities of the controller. However, it is important to note that the methodology presented can be easily extended to design a controller aimed at rejecting a broader range of harmonics. By repeating the provided methodology and incorporating additional integral actions on the desired harmonic components, it is possible to construct a controller that effectively addresses a wider set of harmonics, catering to specific control requirements and system characteristics.
Traditionally, when assessing the harmonic content of an AC signal, two methods are commonly employed, as outlined in the EU 50160:210 standard:
\begin{itemize}
\item assessing the relative magnitude between the $k$-th harmonics and the fundamental component, for instance, standards like D0160 provide specific limits for individual harmonics,
\item computing and minimizing distortion rates, such as the Total Harmonic Distortion (THD), defined as follows:
\begin{equation}\label{THD}
  THD(x) = \sqrt{\frac{\sum_{k \geq 2} |x_k|^2}{|x_1|^2}}.
\end{equation} 
\end{itemize}

Our definition of a phasor aligns perfectly with these considerations for evaluating harmonic content, making it a strong candidate for harmonic content regulation techniques.


\subsection{Measured signals}
The proposed control strategy uses the following signals, which are considered known, measured, or estimated: the three-phase AC source grid-side voltage $e_\abc$, current $i_\abc$, phase $\theta$, pulsation $\omega$, and DC bus voltage $v_\dc$. It is important to note that the measurement of the DC load current $i_\dc$ is not taken into account. Throughout the remainder of the paper, we assume that $\omega$ is constant, and if a Phase-Locked Loop (PLL) is used to estimate it, its dynamics can be neglected.


\subsection{Setpoint and harmonic equilibrium}
In nominal mode, we consider a constant load current $i_\dc$ and assume a sinusoidal input $e_a = -\sqrt{2}E_\mathrm{rms}\cos(\omega t)$. The setpoint for the system can be determined using the $dq$ equations~\eqref{model:converter_detailed_dq_time}. We choose a constant reference value $v_\mathrm{dc_{ref}}$ and impose $i_\mathrm{q}=0$ and a zero voltage sequence, which corresponds to $d_{0}=\frac{1}{3}\sum d_\abc=0.5$. By solving the steady state $dq$ equation, we obtain the equilibrium values $(i_{dq},v_{dc})^e$ and $d_{dq}^e$. Thus, we easily deduce the corresponding setpoint in the $abc$ frame for both the time and harmonic domains using the following relations:
\begin{align}
&V_\mathrm{dc,0}=v_\mathrm{dc,ref} \text{ and }V_\mathrm{dc,k}=0 \text{ for }k\neq 0\\
&I_{a,0}=0,\ I_{a,1}= \frac{1}{\sqrt{6}}\left(i_\mathrm{d} - j i_\mathrm{q}\right),\  I_{a,k}=0 \text{ for } k>1\\
&D_{a,0}= 0.5,\ D_{a,1}= \frac{1}{\sqrt{6}}\left(d_\mathrm{d} - j d_{q}\right),\ D_{a,k}=0 \text{ for } k>1    
\end{align}
and 
$X_b=\hm S_{-\frac{2 \pi}{3}}X_a$,
$X_c=\hm S_{\frac{2 \pi}{3}}X_a$,
where $X$ stands for $I$ or $D$ and where the phase-shift operator is given by $\hm S_\alpha = \mathop{\mathrm{diag}}(e^{j k \alpha},{k \in \mathbb{Z}})$.
Similarly, the nominal harmonic inputs can be expressed as follows:
\begin{align}
E_{a,0}=0, \ E_{a,1} =- \frac{E_\mathrm{rms} \sqrt{2}}{2 },\ E_{k}=0\text{ for } k>1\\
I_\mathrm{dc,0}=i_\dc,I_\mathrm{dc,k}=0\text{ for } k\neq0
\end{align}
It is worth noting that since our defined setpoint represents a $T$-periodic trajectory, it corresponds to an equilibrium $(X^{e}, D^{e},V^{e})$ for the harmonic system~\eqref{model:converter_bilmodel_abc_harmo}. Therefore, it satisfies the following relation:
\begin{equation} 
0 = (\mathcal{A} - \mathcal{N})X^e+ \hm G(X^e)D^e+ \mathcal{B} V^e
\end{equation}

\begin{table}[!t]
    \centering
    \begin{tabular}{|c|c|c|c|}\hline
       Symbol  & Quantities & Value & Unit \\ \hline
       $r$& Phase resistance & 1.15   & $\Omega$  \\ \hline
       $L$ & Phase inductance & 122 & $\mu H$ \\ \hline
       $C$ & Bus Capacitance  &  100 & $\mu F$ \\ \hline
       $R_\mathrm{L}$ & Load nominal resistance & 120 & $\Omega$ \\ \hline
       $f$ & AC frequency & 50 & Hz \\ \hline
       $\omega$ & AC pulsation & 314 & rad/s \\ \hline
       $E$ & AC rms voltage & 45 & V \\ \hline
    $v_\mathrm{dc,ref}$ & DC load nominal voltage & 150 & V \\ \hline
   $ i_\mathrm{dc,n}$ & DC load nominal current &$\frac{v_\mathrm{dc,ref}}{R_\mathrm{L}} = 1.25$ & A \\ \hline
   $ i_\mathrm{sink}$ & Programmable current load & - & A \\ \hline
   $ i_\mathrm{dc}$ & DC load actual current & $i_\mathrm{sink}+\frac{v_\dc}{R_\mathrm{L}}$ & A \\ \hline
   $f_{in,hm}$ & DC load harmonic content frequency & 150 & Hz \\ \hline
    \end{tabular}
    \caption{Experimental Parameters Values}
    \label{tab:numerical_elec_value}
\end{table}

Having defined the control objectives and the setpoint, we now are in position to design and tune a stabilizing control as in~\eqref{control:SF_hmq} and add integral actions as in~\eqref{control:hmq_form}.

\section{Tuning the harmonic model based control}\label{sect:TuningCtrl}

Our objectives encompass several aspects, including the regulation of the mean value of the DC voltage bus, optimization of power factor, and reduction of total harmonic distortion (THD) in the presence of step perturbations and load-induced harmonic content. Furthermore, we delve into saturation strategies to address control signal limitations. To provide a realistic context, the numerical values used in our simulations correspond to those of the experimental setup discussed in the subsequent section (see table \ref{tab:numerical_elec_value}).




\subsection{Stabilizing control} \label{stab_ctrler}
As power converters are known to have stable equilibria due to damping terms, $(\mathcal{A}-\mathcal{A}(\mathcal{D}^e)-\mathcal{N})$ is Hurwitz and Theorem~\ref{stab} can be applied.
Note that its spectrum can be explicitly computed (see for example   \cite{wereleyLinearTimePeriodic1991,riedingerSolvingInfiniteDimensionalHarmonic2022}). 


\paragraph{Tuning $Q$}
$Q(t)$ or, equivalently, $\cal Q$ needs to be carefully selected to penalize the current value on the DC side during voltage transients. The objective is to avoid overcurrent and protect both the power line and the capacitor  ($i_{capa}= C \dot v_\dc \approx d \cdot i$). Additionally, $\cal Q$ should represent a positive definite operator.
To meet these requirements, we choose $Q=\begin{bmatrix}
I_3  & 0 \\ 0 & \alpha
\end{bmatrix}$ a constant matrix. 

Through empirical analysis and experimentation, we have determined that a value of $\alpha =  10^{-4}$  yields satisfactory results for our specific case, particularly during the transient. 
\paragraph{Tuning $ H_1$}
 $H_1$ is selected as a constant scalar value. We find that an appropriate starting value to tune $H_1$ and avoid saturation of the control is $H_1 = \frac{1}{50}\frac{1}{\overline \sigma \left( {\hm G(X^{e})}^* \hm P \right)}=0.613$ where $\overline \sigma$ refers to the maximal singular value. 

 To gain insight into the dynamic behavior of the obtained control, we can analyze the eigenvalues of the harmonic state matrix $\mathcal{A-N}  - \mathcal{G}(X^e)\mathcal{H}_1 \mathcal{G}^*(X^e) \mathcal{P}$ of the linearized harmonic closed-loop model.
 This analysis can be performed using results given by \cite{wereleyLinearTimePeriodic1991,zhouSpectralCharacteristicsEigenvalues2004} or solving explicitly the eigenvalue problem as detailed in \cite{riedingerSolvingInfiniteDimensionalHarmonic2022}.

\subsection{ Integral actions}
In order to enhance robustness against uncertainties or periodic perturbations, two types of harmonic integral actions are incorporated. The first type of integral action is the classical integral action applied to the error signal, aimed at regulating the $0$-phasor component of the relevant signals such as DC voltage and $i_q$. The second type is a harmonic space integrator applied to pairs of conjugate phasors (of order $\pm k, k\in \mathbb{Z}$), with the goal of rejecting undesired harmonic content in the regulated periodic trajectories of $i_\abc$.  We show that the second type of integral action can be expressed as a resonant controller, thereby establishing a compelling connection between our methodology and the family of proportional-resonant (PR) controllers \cite{sritaModelingSimulationDevelopment2022, teodorescuProportionalresonantControllersFilters2006} as well as generalized integrators \cite{yuanStationaryframeGeneralizedIntegrators2002, zhangMethodBasedGeneralized1999}. However, our approach extends beyond these existing frameworks, encompassing a broader context and offering novel insights into the control design.
 
\subsubsection{0-Phasor integral action} \label{int_action}


we incorporate an integral action on both $v_\mathrm{dc}$ and $i_\mathrm{q}$:
\begin{equation} \label{integrator_base_expression_time}
\begin{aligned}
    \dot z_1&= \ell_1 (v_\dc-v_\mathrm{dc}^e)\\
    \dot z_2&= \ell_2  i_\mathrm{q}
\end{aligned}
\end{equation}
where the gains $\ell_{1,2} \in \R $ are parameters useful for adjusting the integrator dynamics.
Equation~\eqref{integrator_base_expression_time} can be reformulated in a canonical form, as expressed in Equation~\eqref{forwarding:OLC_time}, by taking into account:
$L(t) =  \mathop{\mathrm{diag}}(\ell_1, \ell_2)$, $O\equiv0$, and $$C(t) =\begin{bmatrix}
     0 & 1 \\ -\sqrt{\frac{2}{3}}\sin_3(\omega t)   & 0\\
\end{bmatrix}$$ where $\sin_3(\omega t) = \begin{bmatrix} \sin(\omega t)\  \sin(\omega t - \frac{2 \pi }{3})\   \sin(\omega t + \frac{2 \pi }{3}) \\
\end{bmatrix} $.\\
Finally, in the harmonic domain, Equation~\eqref{integrator_base_expression_time} can be written in the canonical form as given by Equation~\eqref{forwarding:OLC_hmq}: 
\begin{equation}\begin{aligned}\label{hm_regul_vdciq}
\begin{bmatrix}
    \dot Z_1 \\ \dot Z_2
\end{bmatrix} =& \begin{bmatrix}- \N &0\\0&-\N\end{bmatrix}\begin{bmatrix}
    \dot Z_1 \\ \dot Z_2
\end{bmatrix} + \hm L \hm C(X-X^{e}) \\ 
\end{aligned}
\end{equation}
with  $ \hm L = \mathop{\mathrm{blkdiag}}(\ell_1 \I , \ell_2 \I)$, $\hm C = \begin{bmatrix} 0 & \I \\ -\sqrt{\frac{2}{3}} \hm S_{in,3}^* & 0\end{bmatrix}$  where $\hm S_{in,3} = \T (\sin_3)$ and $\hm O \equiv 0$.


\subsubsection{Harmonic rejection} \label{hm_action}

Let us show that the integral action can also be used to reject harmonic perturbations on a $k-th$ phasor of a given output $y=Cx$.
To this aim, let us write equation~\eqref{forwarding:OLC_hmq} as follows:
\begin{equation}\dot Z = \begin{bmatrix}
-\mathcal N & -\omega k \mathcal{I} \\ \omega k \mathcal{I}& -\mathcal N
\end{bmatrix} Z + \begin{bmatrix}
\alpha \hm I \\ \beta \hm  I
\end{bmatrix} Y\label{Oh}
\end{equation} 
This structure enables us to incorporate integral action on phasors of order $\pm k$. Assuming asymptotic stability of the harmonic system, we can derive the following relations for the $l$-th lines of the above equations when $\dot Z=0$:
\begin{align}
0 &= -Z_{1,l} j \omega l - Z_{{2,l}} \omega k + \alpha Y_{l}\\
0 &= Z_{1,l} \omega k - Z_{2,l} j \omega l + \beta Y_{l}
\end{align}
Therefore, for $l=\pm k$, multiplying the second line by $j$ and summing leads to:
\begin{equation*}
    0 = (\alpha - j \beta)Y_{\pm k} \text{ e.g. } Y_{\pm k}=0 
\end{equation*}
which proves that a rejection of the $k$-phasor of $Y$ is performed.
In the time domain,~\eqref{Oh} reduces to the classical resonant system: 
 \begin{equation} \label{canonhmqrej}
        \dot z =\begin{bmatrix}
            0 & -k \omega \\ k \omega & 0
        \end{bmatrix}z + \begin{bmatrix}
            \alpha \\ \beta 
        \end{bmatrix} y(t)
        =O_k z +\begin{bmatrix}
            \alpha \\ \beta 
        \end{bmatrix} y(t)
\end{equation}
with $O_k = k \omega R$ the resonant matrix of pulsation $k \omega $ where $R$ is given by~\eqref{model:converter_detailed_dq_time}.

Based on this result and considering the secondary objective of addressing the third-order harmonic injection on the load side, which affects the second and fourth phasors of $i_\abc$ (respectively, the third phasor of $i_\dq$), we focus on the following output:

\begin{equation}\label{C_osc}\begin{bmatrix}
    i_q \\ 
    \overline{i_d}
\end{bmatrix} = C(t)\overline{x}\end{equation}
where  
$$
C(t) =\sqrt{\frac{2}{3}} \begin{bmatrix}
-\sin_3(\omega t) \\ 
\cos_3(\omega t) 
\end{bmatrix}
$$
where $\cos_3(\omega t)=[ \cos(\omega t) \ \cos(\omega t - \frac{2 \pi }{3}) \  \cos(\omega t + \frac{2 \pi }{3})]$.
To achieve rejection of the third harmonic component on $i_\dq$, we introduce the following additional variables:
\begin{align}
    \begin{bmatrix}
        \dot z_3 \\
        \dot z_4 \\
        \dot z_5 \\
        \dot z_6 
    \end{bmatrix} &= \begin{bmatrix}
        O_3 & \\ & O_3 
    \end{bmatrix} \begin{bmatrix}
        \dot z_3 \\
        \dot z_4 \\
        \dot z_5 \\
        \dot z_6 
    \end{bmatrix} + \begin{bmatrix}
        \ell_3 & 0 \\ 0 & 0 \\ 0 & \ell_4 \\ 0 & 0
    \end{bmatrix}   \begin{bmatrix}
    i_q \\ 
    \overline{i_d}
\end{bmatrix}\label{har_rej}
\end{align}


\subsubsection{Final control determination} \label{whole_ctrl}
We are now ready to express the complete set of integral actions that fulfill both the primary and secondary objectives. This entails considering both $(z_1, z_2)$ and $(z_3, z_4, z_5, z_6)$, which are given by Equations~\eqref{integrator_base_expression_time} and~\eqref{har_rej}, respectively.
Combining these equations to obtain the canonical form, as presented in Equation~\eqref{forwarding:OLC_time} in the time domain and Equation~\eqref{forwarding:OLC_hmq} in the harmonic domain, is straightforward when considering a single output given by:
$$y=C(t)(x-x^e)= \begin{bmatrix}
    v_\dc - v_\dc^e \\ i_q - i_q^e \\ i_d-i_d^e 
\end{bmatrix}= \begin{bmatrix}
    v_\dc - v_\dc^e \\ i_q \\ i_d-i_d^e 
\end{bmatrix}.$$
The resulting matrices for Equation~\eqref{forwarding:OLC_time} are as follows:
\begin{equation}\label{finalOLC_time}
\resizebox{.8\hsize}{!}{$
O = \begin{bmatrix}
0 \\ 
& 0 \\ 
&&0&-3 \omega \\ 
&& 3\omega&0 \\ 
&&&& 0&-3 \omega \\ 
&&&& 3\omega &0
\end{bmatrix}\ L = \begin{bmatrix}
\ell_1&0&0 \\ 0& \ell_2&0 \\ 0&\ell_3&0 \\0&0&0 \\ 0&0&\ell_4 \\ 0&0&0
\end{bmatrix}$}
\end{equation}
and for~\eqref{forwarding:OLC_hmq}:
$\hm O = O \otimes \I, \hm L = L \otimes \I$, and $$\hm C = \begin{bmatrix}
    0 & \I  \\ -\sqrt{\frac{2}{3}}\hm S_{in,3} & 0 \\  \sqrt{\frac{2}{3}}\hm C_{os,3} & 0 
\end{bmatrix}$$  with $\hm C_{os,3} = \T (\cos_3)$


\begin{figure}[!t]
\centering
\includegraphics[width=0.95\columnwidth]{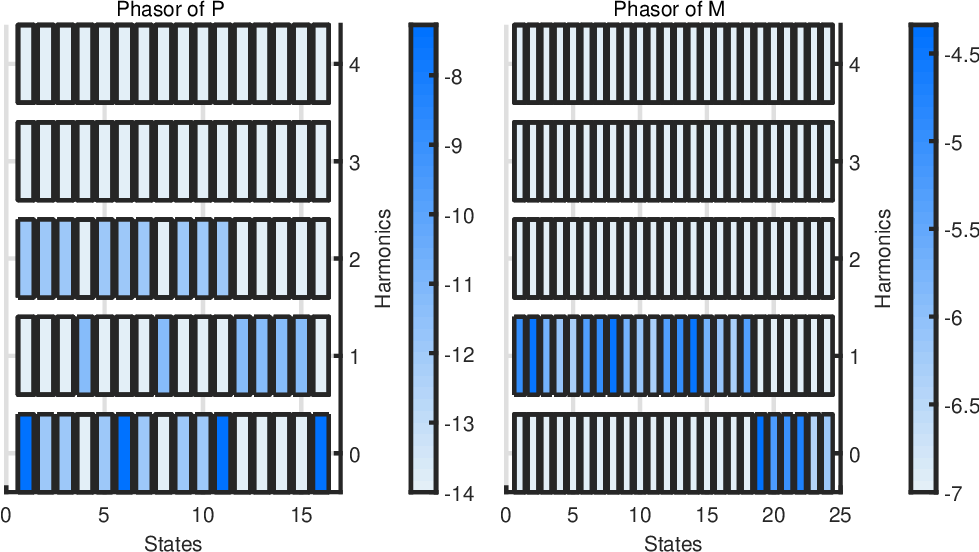}\centering
\caption[Harmonic content after solving]{Harmonic content of $\mathrm{col}{P}, \mathrm{col}{M}$, log scale}
\label{fig:2_hmq_content_PM}
\end{figure}

Finally, in the harmonic domain, we solve the associated Sylvester equation~\eqref{Sylvester:hm} to determine~\eqref{control:hmq_form}, from which~\eqref{control:time_form} is deduced using~\eqref{inv_four_reconstruc}.
Furthermore, using trial and error to achieve satisfactory performances, we chose the following parameters: 
$\ell_1=0.1$, $\ell_2=\sqrt{\frac{2}{3}}$ and  $\ell_3=\ell_4=0.14\sqrt{\frac{2}{3}}$. 
Aggressiveness of the control is tuned via $H_2 = \alpha\operatorname{blkdiag}(1,0.1,I_{d,2},I_{d,2}) $ in~\eqref{control:time_form} where $\alpha = \frac{1}{H_1} \frac{1}{50}\frac{1}{\overline{\sigma}(\hm G(X^e) \hm M^* \hm M)} =6.919$ and $\overline{\sigma}(y)$ represents the maximum singular value of $y$.

In practice, the infinite-dimensional Lyapunov harmonic equation~\eqref{lyapunov:hm} and the Sylvester equation~\eqref{Sylvester:hm} are solved using the algorithm developed in \cite{riedingerSolvingInfiniteDimensionalHarmonic2022} and \cite{riedingerHarmonicPolePlacement2022} respectively. The truncation level required in these algorithms has been set at ten harmonics, which proves more than sufficient to determine an accurate solution. The total computation time is less than $0.05$ sec. Finally, in order to visualize the harmonic contents of each component of matrices $P$ and $M$, we denote $\operatorname{col}(X)$ as the vectorization of a matrix $X$, which is formed by stacking the columns of $X$ into a single column vector. Figure~\ref{fig:2_hmq_content_PM} displays the modulus of each harmonic component of $\operatorname{col}(P)$ and $\operatorname{col}(M)$. It is evident that the harmonic content of both $M$ and $P$ is significantly reduced.

\subsection{Taking into account control saturations}
Physical switches in the system limit duty cycle values to the range $[0, 1]$. However, selecting a small enough gain value to prevent saturation does not guarantee that the control will always stay within these bounds, especially in the presence of load perturbations.

In the previous section, we introduced the control~\eqref{control:time_form} without considering the necessary saturation of its values in the range $[0, 1]^3$. It is crucial to prove that the presented control is compatible with these saturation constraints while ensuring system stability. In this section, we address this by introducing a saturation method inspired by \cite{beneuxAdaptiveStabilizationSwitched2019}. This method incorporates control saturation while preserving global stabilization properties.

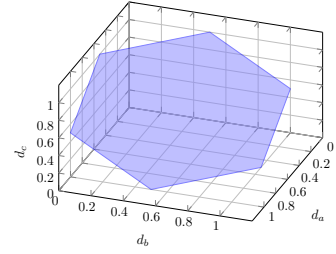
\begin{figure}[!t]
\centering\resizebox{0.5\columnwidth}{!}{
\begin{tikzpicture}
     \begin{axis}[
        xlabel=$d_a$,
        ylabel=$d_b$,
        zlabel=$d_c$,
        view={110}{40},
        xmin=0, xmax=1.2,
        ymin=0, ymax=1.2,
        zmin=0, zmax=1.2,
        ztick={0,0.2,...,1},
        ytick={0,0.2,...,1},
        xtick={0,0.2,...,1},
        tick style={grid=major},
        grid=major
    ]
    \addplot3[
        color=blue,
        fill=blue!50,
        opacity=0.5,
    ]
   coordinates {
    (0,1,0.5)
    (0,0.5,1)
    (0.5,0,1)
    (1,0,0.5)
    (1,0.5,0)
    (0.5,1,0)
    (0,1,0.5) 
    };
    \end{axis}
\end{tikzpicture}}
\caption{Convex polygon: $d_a+d_b+d_c=1.5$ for $0\leq d_\abc\leq 1$}
\label{fig:3_parametric_surface}
\end{figure}

To establish that the control signal $d(t)$ obtained from~\eqref{control:time_form} belongs to the domain defined by $\mathfrak{D}: d_1 + d_2 + d_3 =1.5$ at every time $t$, we can examine the expression of $G(x)^T$ which is given by:
$G(x)^T = \begin{bmatrix}
\frac{C_{33}}{L} v_\dc & \frac{i_\abc}{C}
\end{bmatrix}$.
Since the sum of the rows of $G(x)^T$ is always equal to zero, by inspecting~\eqref{control:time_form}, we can observe that $\sum d_\abc = \sum d^e = 1.5$ in accordance with the defined setpoint in Section III.B. 
Therefore, following the approach presented in \cite{beneuxAdaptiveStabilizationSwitched2019}, the control $d_\abc$ given by~\eqref{control:time_form} can be saturated by projecting it onto the surface of the convex polygon defined by $\mathfrak{D}: d_1 + d_2 + d_3 = 1.5$ with the constraint $d_\abc \in [0\ 1]^3$ (see Fig. \ref{fig:3_parametric_surface}).

If we define $\delta d$ as $\delta d = - H_1 G^T(x) [P(x-x^{e}) - M^T H_2 (z- M(x-x^{e}))$ in~\eqref{control:time_form}, then the saturation of the control signal can be achieved using the following expression:
\begin{equation}
\sat(d) = d^{e} + \min_{i=a,b,c}(\alpha_i) \delta d 
\end{equation}
where \begin{equation}
\alpha_i = \begin{cases}
\min\left(1, \frac{1-d^{e}_i}{\delta d_{i}}\right) & \mathrm{if} \; \delta d_{i} >0 \\ 
\min\left(1, \frac{-d^{e}_i}{\delta d_{i}}\right) & \mathrm{if} \;\delta d_{i}<0 \\ 
1 & \mathrm{if}\; \delta d_{i}=0 \\ 
\end{cases}
\end{equation}
This saturation ensures global stability (see the proof in \cite{blinNecessarySufficientConditions2022} Proposition 9). 
Notice that the integral formulation allows us to introduce a deviation from reference term $\delta r(t)$ in the control law \ref{forwarding:OLC_time}. Therefore, our final control law takes the following form:
\begin{equation} \label{forwardingwithref}
\begin{aligned}
d =& d^{e} + \alpha(t) \delta d(x-x^{e},z)\\ 
\delta d =& -H_1(t)G(x)^T P(t)(x-x^{e})\\
&+H_1(t)G(x)^T M^T(t) H_2(t) (z - M(t) (x-x^{e}))\\
\dot z =& Oz + {L(C(t)}(x-x^{e})+\delta r(t))
\end{aligned}
\end{equation}

\subsection{Simulation and validation}
\begin{figure}[!t]
\centering
\includegraphics[width=\columnwidth]{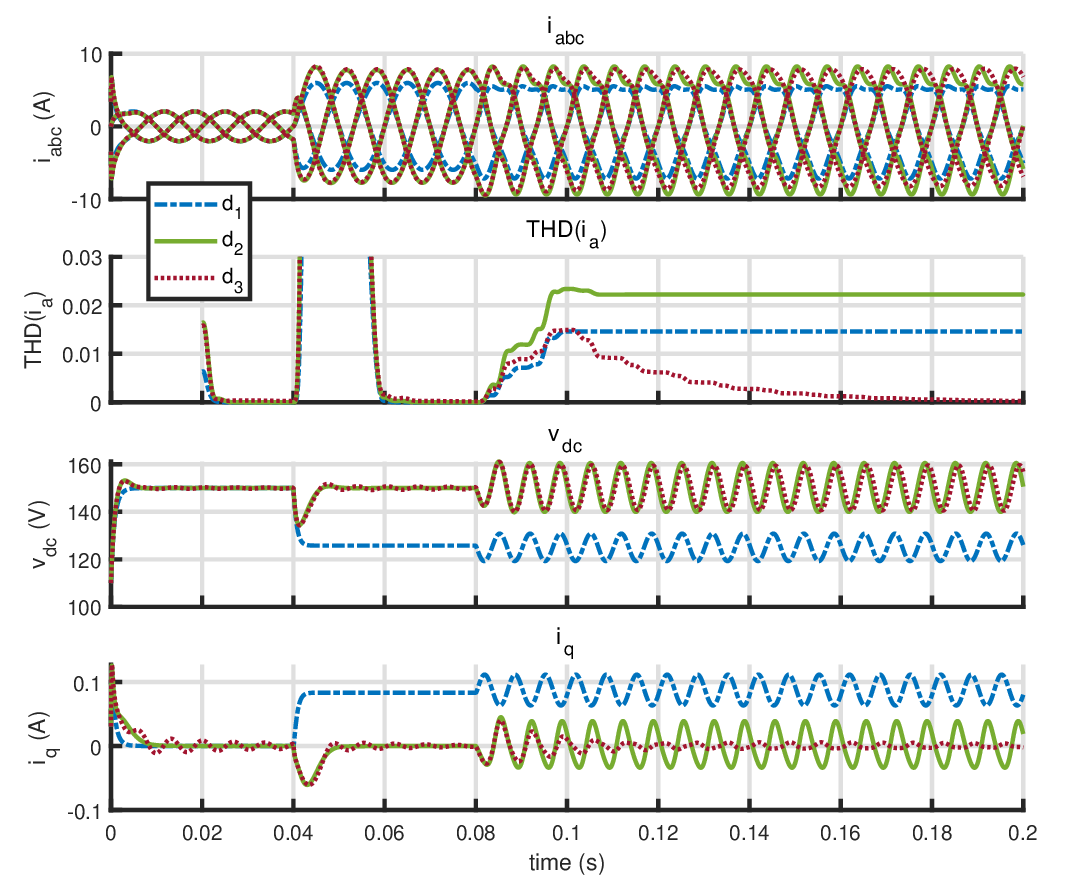}
\centering
\caption{$i_\abc$, $\mathrm{THD}(i_a)$, $v_\dc$ and $i_q$ responses to a step and $3$rd harmonic injection in $i_\dc$ for control laws : $d_1$, $d_2$, $d_3$.}
\label{fig:compar_start_simu}
\end{figure}

Simulation results, performed using MATLAB, are presented in Figure~\ref{fig:compar_start_simu}. The simulation starts from an initial condition corresponding to the three-phase diode rectifier mode. At time $t=2T=0.04 \mathrm{s}$, a step load current of $\delta i_\dc=3$A is applied, followed by a perturbation injection of a $150$ Hz ($3$rd order phasor) load current with an amplitude of 1A at $t=4T=0.08 \mathrm{s}$. To illustrate the control design, three controllers are compared:
\begin{itemize}
\item[$d_1$:] a simple stabilizing controller in the form given by \eqref{control:stabilizing_time}, tuned as described in Section \ref{stab_ctrler}.
\item[$d_2$:] a forwarding controller in the form of \eqref{control:time_form}, including integral actions only on the 0-phasor of $v_\dc$ and $i_\dq$, tuned as explained in Section \ref{int_action}.
\item[$d_3$:] the full integral action controller \eqref{control:time_form} designed and tuned as described in Section \ref{whole_ctrl}, aimed at rejecting the $2$nd and $4$th harmonic components in $i_\abc$.
\end{itemize}
All controllers successfully reach the set point under nominal conditions. At $t=2T$, only the controllers with integral action are able to regulate the DC bus voltage value (and $i_q$). Among the three controllers, only the third controller achieves harmonic rejection in the current phase, as demonstrated by the Total Harmonic Distortion (THD) of $i_a$.

\begin{figure}[!ht]
\centering
\includegraphics[width=0.95\columnwidth]{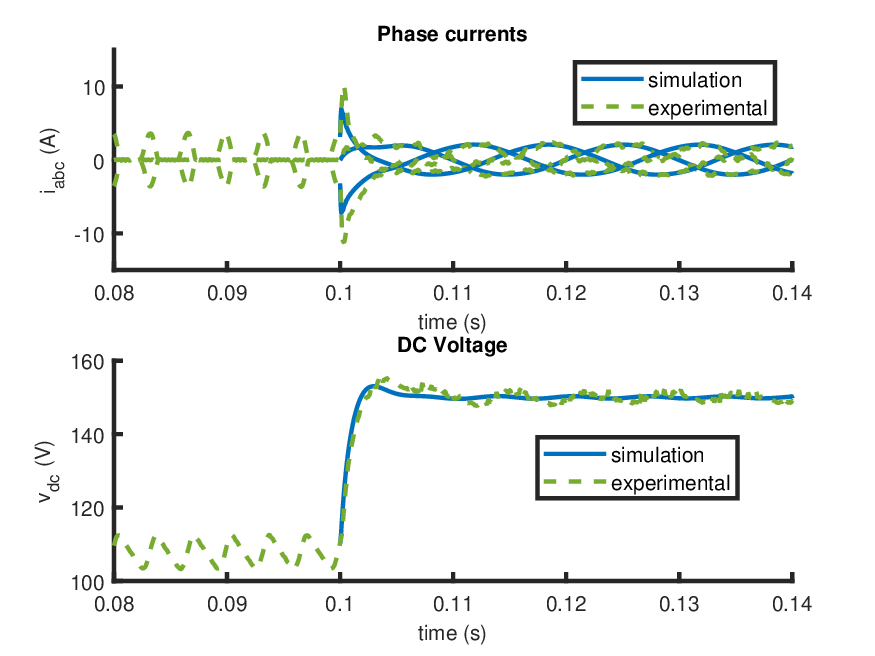}\centering
\caption{Comparison between simulation and experimental acquisition, startup to nominal set-point}
\label{fig:compar_start_SimuvExp}
\end{figure}

Figure~\ref{fig:compar_start_SimuvExp} showcases simulations of the proposed control ($d_3$, tuned in Section \ref{whole_ctrl}) applied to the startup of the AC/DC converter, along with the actual experimental implementation and data acquisition from the testbed, which will be presented in the following section. The figure showcases the remarkable predictability and accuracy of the control. The close agreement between the simulated results and the data obtained from the real implementation strengthens the effectiveness and reliability of the control strategy.

\section{Practical implementation}\label{sect:PractImpl}
To validate the performance and effectiveness of our control algorithm in a practical, real-world scenario, we conducted practical testing using a setup consisting of a 3-phases grid connected to a controlled DC load. This setup provided us with the opportunity to assess the control's performance under realistic conditions and verify its suitability for practical applications. 

\subsection{Experimental setup}
\subsubsection{Test bench}
\paragraph{Setup}

A 3-phase grid is replicated using a \href{https://assets.testequity.com/te1/Documents/pdf/61700.pdf}{Chroma 61704 controllable three-phase power supply} connected to \href{https://magnetic-components.mpsind.com/Asset/HWI57A2-071A-121P_A.pdf}{Line inductance (122µH)} and \href{https://docs.rs-online.com/a583/0900766b81544122.pdf}{3-phase 2-level voltage inverter}. The control is implemented using \href{http://www.dttechsolutions.com/images/products/DSpace/dSPACE-MicroLabBox_Product-Brochure_2020-01_EN.pdf}{MicrolabBox} and \textsc{Simulink}. A \href{https://www.regatron.com/assets/resources/Documents/Technical-Datasheets/TC.GSS/2210_TC.GSS.20.600.4WR.S_datasheet_EN_20200423.pdf}{TopCon Reversible DC Power Supply} is connected on the DC side of the converter in parallele to a 120 $\Omega$ tunable resistance. The DC power supply is controlled through the Dspace MicroLabBox software to act as a programmable current load. The high bandwidth of the TopCon allows for injecting chosen harmonics.

\begin{figure}[!t]
\centering
\includegraphics[width=0.95\columnwidth]{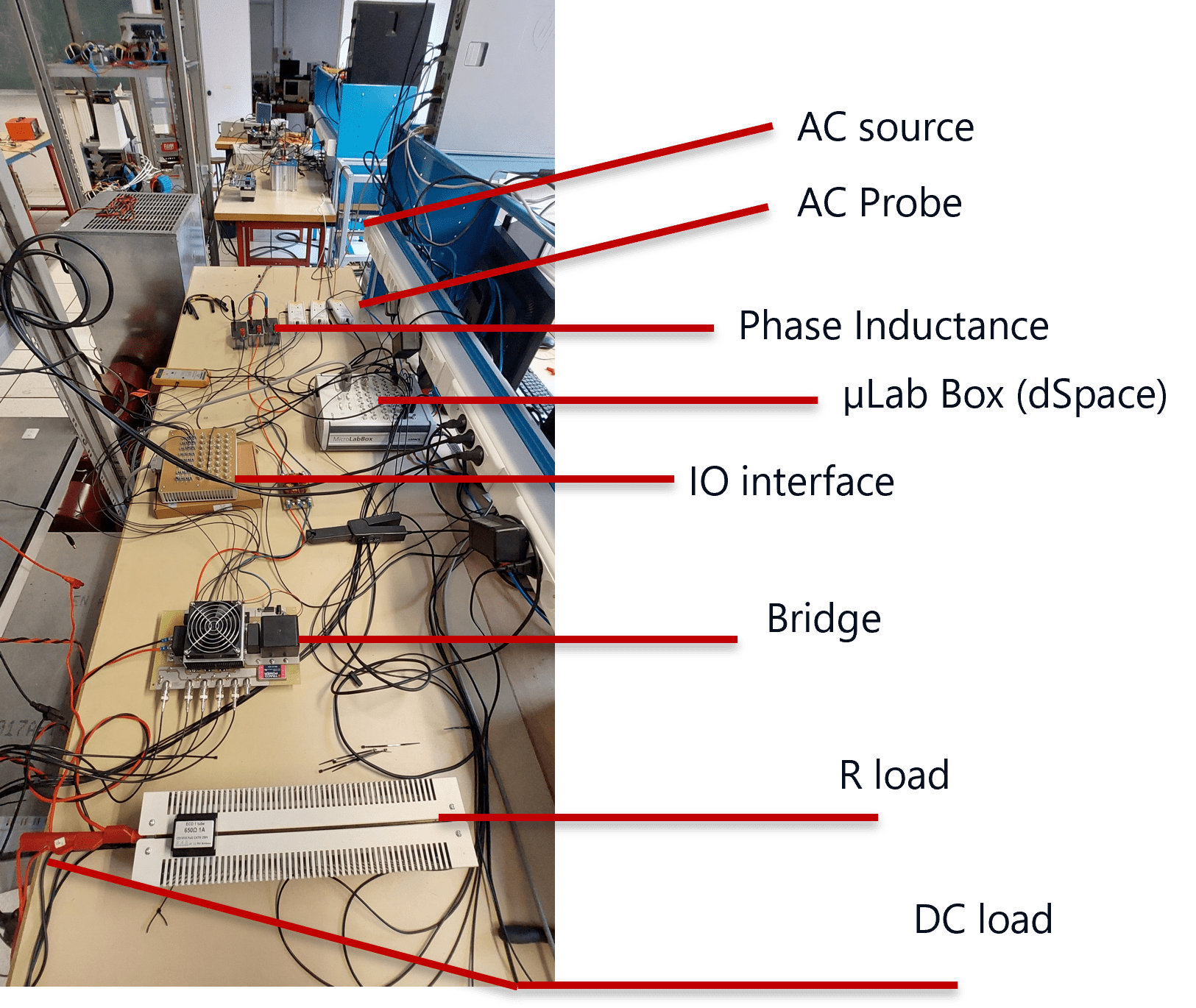}\centering
\caption{Test bench, DC and AC source are outside picture}
\label{fig:simu_tst_bench}
\end{figure}

\paragraph{Measurement}

\href{https://www.es-france.com/index.php?controller=attachment&id_attachment=11329}{Testec TT SI 9001 tension probes} are used for direct measurements of AC voltage relative to the neutral point from the AC source. Additionally, measurements of phase-to-phase voltage, DC voltage, capacitor current, and AC phase current are performed. These measurements provide comprehensive data on the system's electrical variables for real-time analysis of the control algorithm's performance.

\subsubsection{Setup, discussion}

In contrast to ideal simulation conditions, the AC and DC sources in the experimental setup exhibit non-ideal characteristics. The phase voltages are not purely sinusoidal, and reconstruction of electric angle and pulsation using a phase-locked loop (PLL) technique is required. The DC load current source, controlled by the MicroLabBox, closely follows the reference signal. However, nonlinearities and imperfections introduce harmonic components that are multiples of the reference frequency.

\subsection{Control implementation}
The control is implemented in discrete time at a frequency of 20 kHz to remain in industrial standard.

\subsubsection{Periodic matrices}

As the electric angle is easily computed using a PLL, the matrix expression of $P$ (and $M$) will be shifted from time/frequency dependency ($P(t) = \sum_{k\in \Z} P_k e^{jk \omega t}$) to a more appropriate phase angle description as: $P(t) = \sum_{k\in \Z} P_k e^{jk \theta(t)}$. 
As explained previously, $P$ and $M$  can be truncated to include only low-order harmonics as their coefficients rapidly decrease to values that are indistinguishable from measurement noise (see Figure \ref{fig:2_hmq_content_PM}). 
Finally, in order to only manipulate real gains, $P$ is expressed as a sum of cosine and sine terms instead of the previous description. Namely, \begin{equation}
    P = P_0+\sum_{k=1}^{h_\mathrm{trunc}} P_{c,k} \cos(k\theta(t)) + P_{s,k}sin(k\theta(t))
\end{equation}  with $h_\mathrm{trunc}=3$
\begin{align}
P_{c,k} &= 2 \operatorname{Re}(P_{k})\\
P_{s,k} &= -2 \operatorname{Im}(P_{k})
\end{align}

\subsubsection{Phase Locked Loop}
In order to determine the phase of the system for the implementation described earlier, a standard phase-locked loop (PLL) is utilized. The equations governing the PLL can be expressed as follows:
\begin{equation}\label{pll}
H_{PLL} = 1E5 \frac{1 + s/(5\times2\pi)}{1+s/(500\times 2\pi)} \frac{1}{s^2}
\end{equation}

\subsubsection{Discretization of the harmonic control}
Finally, as the control is sampled at a frequency of 20 kHz, the final form of the control is given by:
\begin{align}
d_{kT_s} =& d^{e}(\hat \theta_{kT_s}) \\
&- H_1 G(x_{kT_s})^T P(\hat\theta_{kT_s})(x_{kT_s}-x^e(\hat\theta_{kT_s})) \nonumber\\
&+ H_1 G(x_{kT_s})^T M(\hat\theta(t))^{T}H_2 M(\hat\theta_{kT_s})\nonumber\\&\left[z_{kT_s} - M(\hat\theta_{kT_s})(x_{kT_s}-x^{e}(\hat\theta_{kT_s}))\right]\nonumber \\
z_{(k+1)T_s} =& O_\mathrm{d} z_{kT_s} + L_\mathrm{d} C(\hat\theta_{kT_s})(x_{kT_s} - x^{e}(\hat\theta_{kT_s})) 
\end{align}
with $O_\mathrm{d}$ representing the discretized block diagonal of the integrator and oscillator matrix. Specifically, in the case of:
\begin{itemize}
\item an integrator $O_\mathrm{d} = 0$, $L_\mathrm{d}=L$, 
\item an oscillator $O_k=\begin{bmatrix}
0 &-k\w \\ k \w & 0
\end{bmatrix}=k\w R$ then 
\begin{itemize}
\item $O_\mathrm{d}= R \sin(kT_s \w)+ I_2 \cos(kT_s \w) $
\item $L_\mathrm{d} = -\frac{1}{k \w}R(O_\mathrm{d}-I_2)L $
\end{itemize}
\end{itemize}
$O_{d}$ can be computed offline for a given pulsation, or fed with $\hat \w$ estimated from the PLL

\section{Experimental performance assessment}\label{sect:ExpAss}

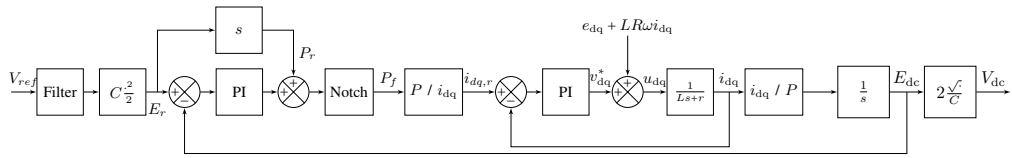
\begin{figure*}[!ht]
\centering \scalebox{0.65}{
\begin{tikzpicture}
\begin{small}
    \sbEntree{Vref};
    \sbBloc{reff}{Filter}{Vref};
    \sbRelier[$V_{ref}$]{Vref}{reff};
    \sbBloc[1]{V2E}{$C\frac{\cdot^2}{2}$}{reff};
    \sbRelier[]{reff}{V2E};
    \sbComp{comp}{V2E};
    \sbRelier[]{V2E}{comp};
    \node[below of=V2E-comp,node distance=1.5em]{$E_r$};
    \sbBloc[1]{reg}{PI}{comp};
    \sbRelier[]{comp}{reg};
    \sbSumh[3.5]{comp2}{reg};
    \sbRelier[]{reg}{comp2};
    \sbBlocL[1]{notchii}{Notch}{comp2}
        \sbDecaleNoeudy[-4]{comp2}{u} ;
        \sbBlocr{r1}{$s$}{u};
        \sbRelieryx{V2E-comp}{r1};
        \sbRelierxy[$P_{r}$]{r1}{comp2};
        \sbBloc{P2iq}{$P$ / $i_\dq$}{notchii};
        \sbRelier[$P_f$]{notchii}{P2iq};
        \sbComp[5]{compi}{P2iq};
        \sbRelier[$i_{dq,r}$]{P2iq}{compi};
        \sbBlocL[1]{PIi}{PI}{compi};
        \sbSumh[4]{compu}{PIi};
        \sbRelier[$v_\dq^*$]{PIi}{compu};
        \sbBloc[1.5]{Hi}{$\frac{1}{Ls+r}$}{compu};
        \sbRelier[$u_\dq$]{compu}{Hi};
        \sbDecaleNoeudy[-4]{compu}{e} ;
        \sbRelier[]{e}{compu};
        \sbNomLien[0.2]{e}{$e_\dq + L R \omega i_\dq$}
        \sbBloc{idq2P}{ $i_\dq$ / $P$}{Hi};
        \sbRelier[$i_\dq$]{Hi}{idq2P};
        \sbRenvoi{Hi-idq2P}{compi}{};
    \end{small}
    \sbBloc{Hv}{$\frac{1}{s}$}{idq2P};
    \sbRelier[]{idq2P}{Hv};
    \sbBloc{Hvv}{$2\frac{\sqrt{\cdot}}{C}$}{Hv};
    \sbSortie{S}{Hvv};
    \sbRelier[$E_\dc$]{Hv}{Hvv};
    \sbRelier[$V_\dc$]{Hvv}{S};
    \sbRenvoi{Hv-Hvv}{comp}{};
 \end{tikzpicture} }
 \caption{Closed loop with PI control, augmented with Notch filter}
 \label{fig:PI_tikz}
\end{figure*}

In order to evaluate the performance of the proposed harmonic control strategy, a comparative analysis is conducted by implementing a PI-type controller (Fig. \ref{fig:PI_tikz}). This comparison serves as a valuable tool to assess and contrast the performance of the control system under different control schemes. The PI controller is implemented by successively tuning the internal current loop, the external voltage loop and the reference filter, while maintaining a factor of 10 between the cutoff frequency of each closed loop.

The control scheme is represented Figure~\ref{fig:PI_tikz}. 
We provide the tuned gains of each PI controller. 
\begin{itemize}
    \item PI current loop :  $K_{P,i}= L \times 6280$ and $K_{I,i}= K_{P,i} \times 6280$, final bandwidth of closed loop current is given by $\omega_i= 2 \pi \times 997 \mathrm{rad/s}$.
    \item PI voltage loop : $K_{P,v}= 2 \times0.707 \times627$,$K_{P,i}= K_{P,i}\times 627$, final bandwidth of closed loop voltage is $\omega_v= 2 \pi \times 99.8 \mathrm{rad/s} $.
    \item Reference filter: $\omega_\mathrm{cons}$  is taken as  $\omega_\mathrm{cons} = \frac{\omega_v}{10} = 62 \mathrm{rad/s}$
\end{itemize}

This particular control scheme has a tendency to amplify the harmonic content existing on the DC bus and propagate it to the AC current. To address this issue, one possible approach is to either decrease the bandwidth of the controller or introduce a notch filter specifically targeting the frequency of concern. However, it is important to note that both of these techniques come with certain trade-offs, particularly in terms of overall system performance. In our case, we will be implementing the latter option, i.e., incorporating a notch filter, to mitigate the undesired harmonic behavior.

Here, we will present 3 experimental situations:
\begin{enumerate}
    \item Transient from three-phase diode rectifier mode toward three-phase controlled voltage rectifier mode (Figure~\ref{fig:ExpAss_start});
    \item Response to a step current input (Figure~\ref{fig:ExpAss_I_Step});
    \item Response to a step load current with an additional 150 Hz harmonic component as stated in the control objectives in Section \ref{whole_ctrl} (Figures \ref{fig:ExpAss_I_step_h3_DCside} and \ref{fig:ExpAss_I_step_h3_ACside}).
\end{enumerate}


Through the presentation of the experimental results, we present time domain acquisition. For AC signals ($i_\abc, d_\abc$), THD calculation over time is performed using $$THD_x (t) = \sum_{k=2}^{k=25} \sqrt{\frac{\vert X_k(t) \vert^2}{\vert X_2(t) \vert^2}}.$$ For DC signals, we compute an analogous Harmonic content quantity $$HC_x (t) = \sum_{k=1}^{k=25} \sqrt{\vert X_k(t) \vert^2},$$ where $X_k(t)$ is the $k$-th phasor of $x$. In all cases, if necessary, dynamics over time of the most relevant phasors are presented. 

\subsection{Start}
\begin{figure}[!t]
\centering
\includegraphics[width=1\columnwidth]{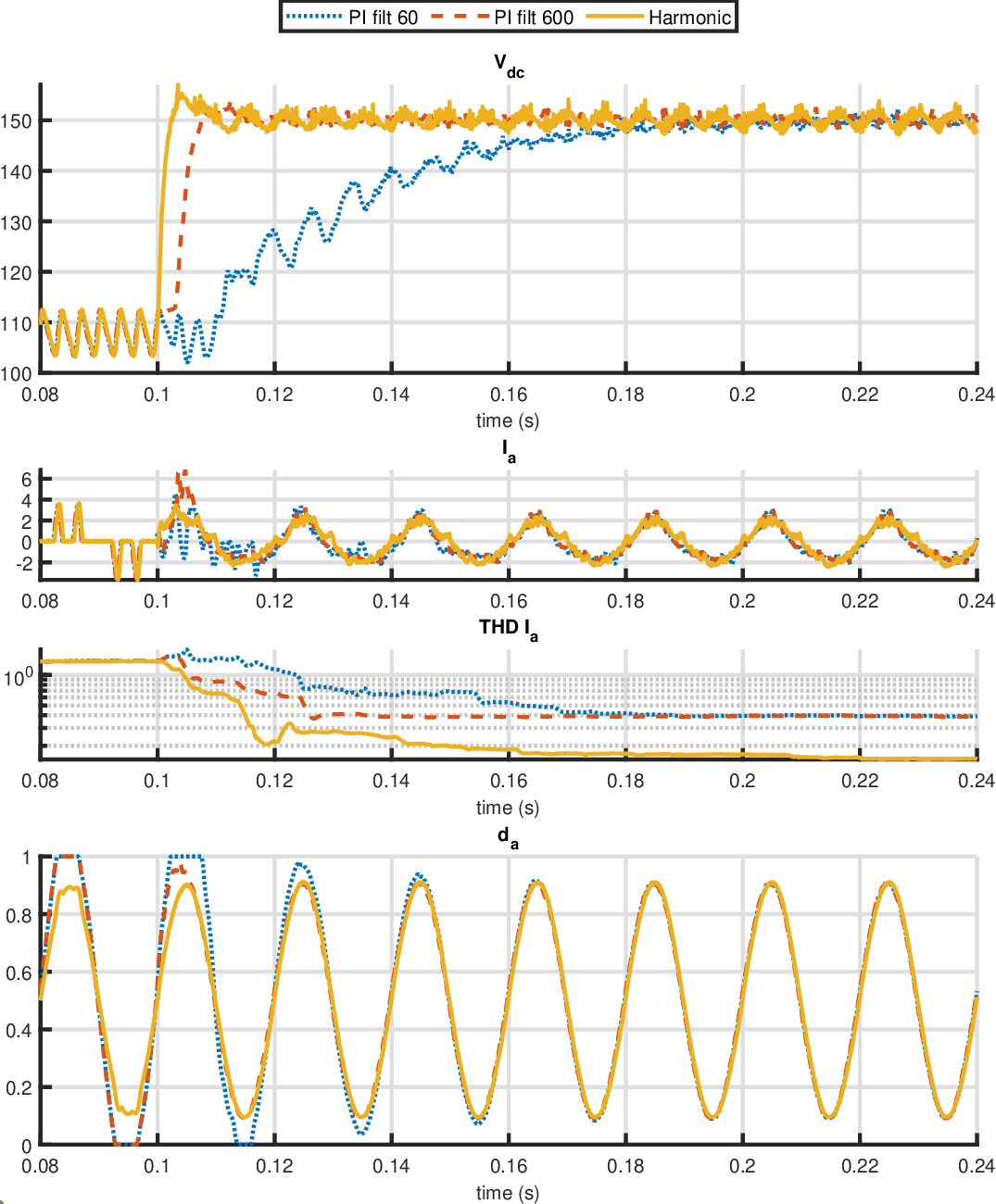}\centering
\caption{Transcient from three-phase diode rectifier mode to reference setpoint : $v_\dc$, $i_a$ (time and THD), $d_a$. Comparison between harmonic based control and two PI controllers with two different reference filters}
\label{fig:ExpAss_start}
\end{figure}

For the experimental validation, the system was initially started with an AC voltage of 45V rms. 
Figure~\ref{fig:ExpAss_start} illustrates the transient response from three-phase diode rectifier mode to controlled mode, towards a DC bus voltage reference of $150V_\dc$. Figure~\ref{fig:ExpAss_start} is divided into three distinct parts for better clarity and analysis: 1) $V_\dc$, 2) $I_a$ and its THD evolution during the transcient, 3) the control applied on the first phase $d_a$.

A second-order filter is added to both PI controllers to smooth the reference signals and slow down the dynamics. The passband frequencies chosen for these filters are 60 Hz and 600 Hz, respectively. It is important to note that the harmonic-based control approach presented in Section \ref{whole_ctrl} does not require the introduction of such a filter.  The inherent nature of the harmonic-based control methodology eliminates the need for additional filtering, as it inherently addresses the harmonic content and stability considerations without the requirement of external filtering techniques.

Figure~\ref{fig:ExpAss_start} provides clear evidence of the superior dynamic performance offered by the harmonic control approach, accompanied by significantly lower harmonic distortion on the current phase as expected. An additional noteworthy observation is that during transients, the PI loop tends to saturate, while our proposed control strategy achieves rapid regulation without any saturation issues. This highlights the  effectiveness of the harmonic control approach in achieving fast and stable regulation, even under challenging transient conditions.

\subsection{Constant perturbation rejection}
\begin{figure}[!t]
\centering
\includegraphics[width=1\columnwidth]{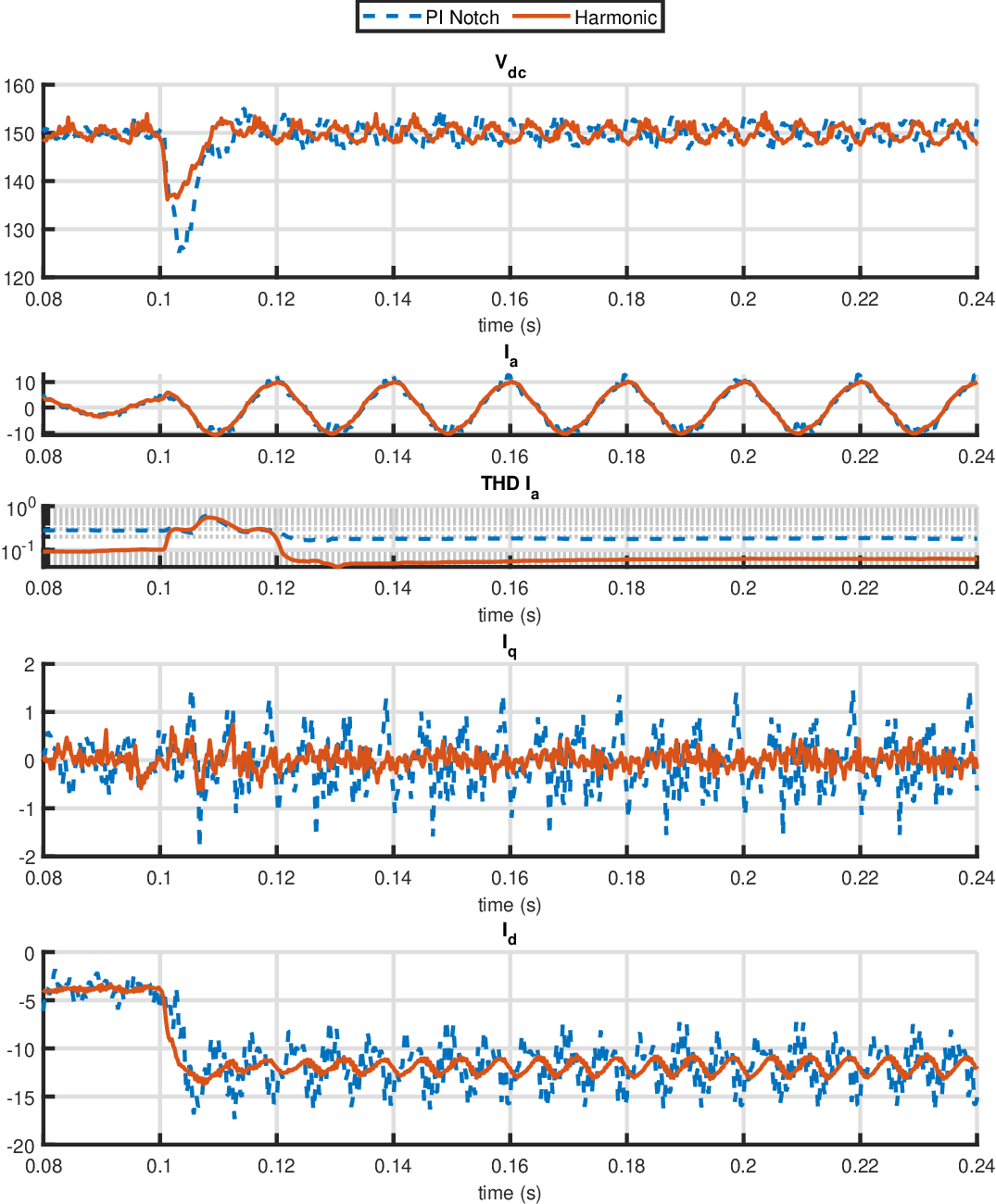}\centering
\caption{Step input on $I_\mathrm{load}$, signals $v_\dc$, $i_a$, $i_\dq$, comparison between harmonic control and PI + Notch strategy}
\label{fig:ExpAss_I_Step}
\end{figure}

During the experimental validation, the system initially operates with a constant current deviation of 1A from the nominal setpoint. From this new equilibrium point, a step change in the current load is introduced, resulting in a 4A variation. The system's behavior during this test is depicted in Figure~\ref{fig:ExpAss_I_Step}. It should be noted that due to non ideal performances of the DC active load current controller, there exists some inherent 3rd harmonic content in the load current, even when a constant reference is applied. As illustrated in Figure~\ref{fig:ExpAss_I_Step}, the implementation of a PI-type control scheme leads to significantly higher harmonic content observed in both $I_a$ and $i_\mathrm{dq}$, along with comparatively lower performance in rejecting DC voltage disturbances when compared to the harmonic control approach.

\subsection{Harmonic rejection}

\begin{figure*}
\centering
\subfloat[]{\includegraphics[width=1\columnwidth]{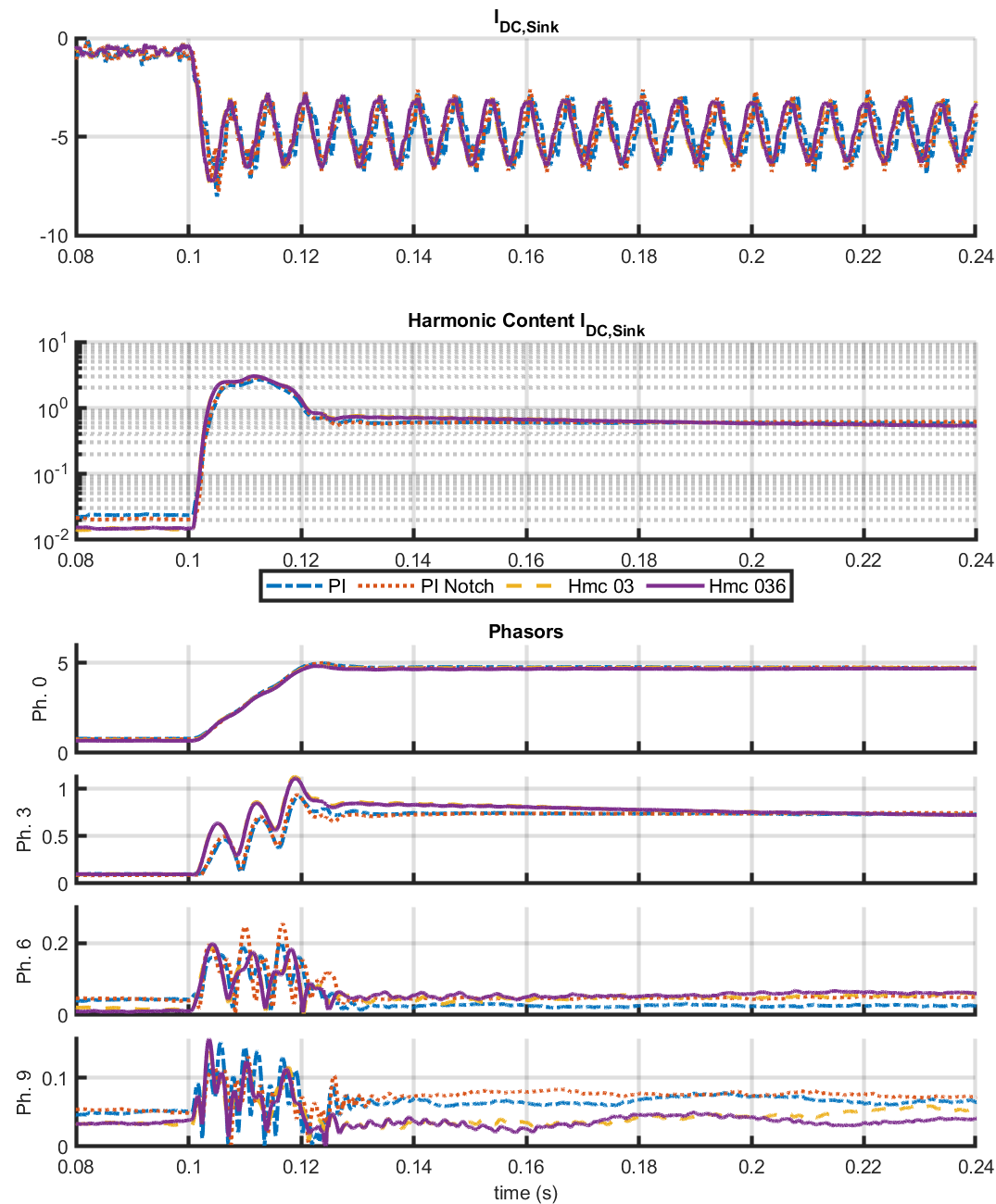}%
\label{fig:ExpAss_I_step_h3_subfig:idcsink}}
\hfil
\subfloat[]{\includegraphics[width=1\columnwidth]{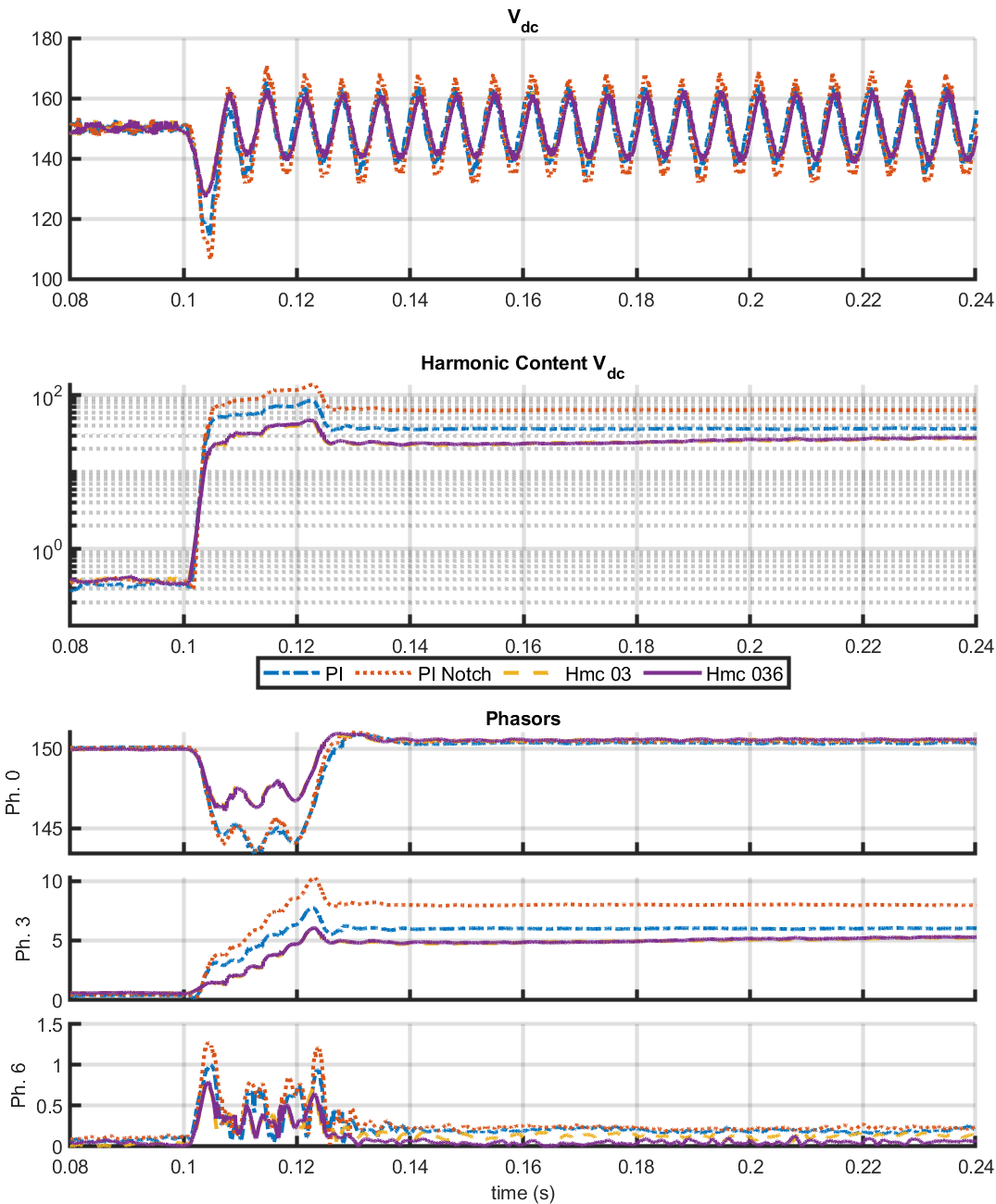}%
\label{fig:ExpAss_I_step_h3_subfig:vdc}}
\caption{Load side harmonic injection: DC BUS (a) $I_{\dc,sink}$, (b) $V_\dc$}
\label{fig:ExpAss_I_step_h3_DCside}
\end{figure*}

{Finally, in the last phase of experimental validation, as previously, we initiate the system by establishing a steady state with a constant current deviation of 1A from the nominal setpoint and we apply a step current to the load side additionned with a third harmonic (150 Hz) component (4A DC, 1A rms at 150 Hz, see Figure~\ref{fig:ExpAss_I_step_h3_subfig:idcsink}) and its multiples. The objective of this test is to evaluate the system's response to harmonic load current disturbances and its capability to mitigate their impact on the AC current.}

\begin{figure*}[!t]
\subfloat[]{\includegraphics[width=1\columnwidth]{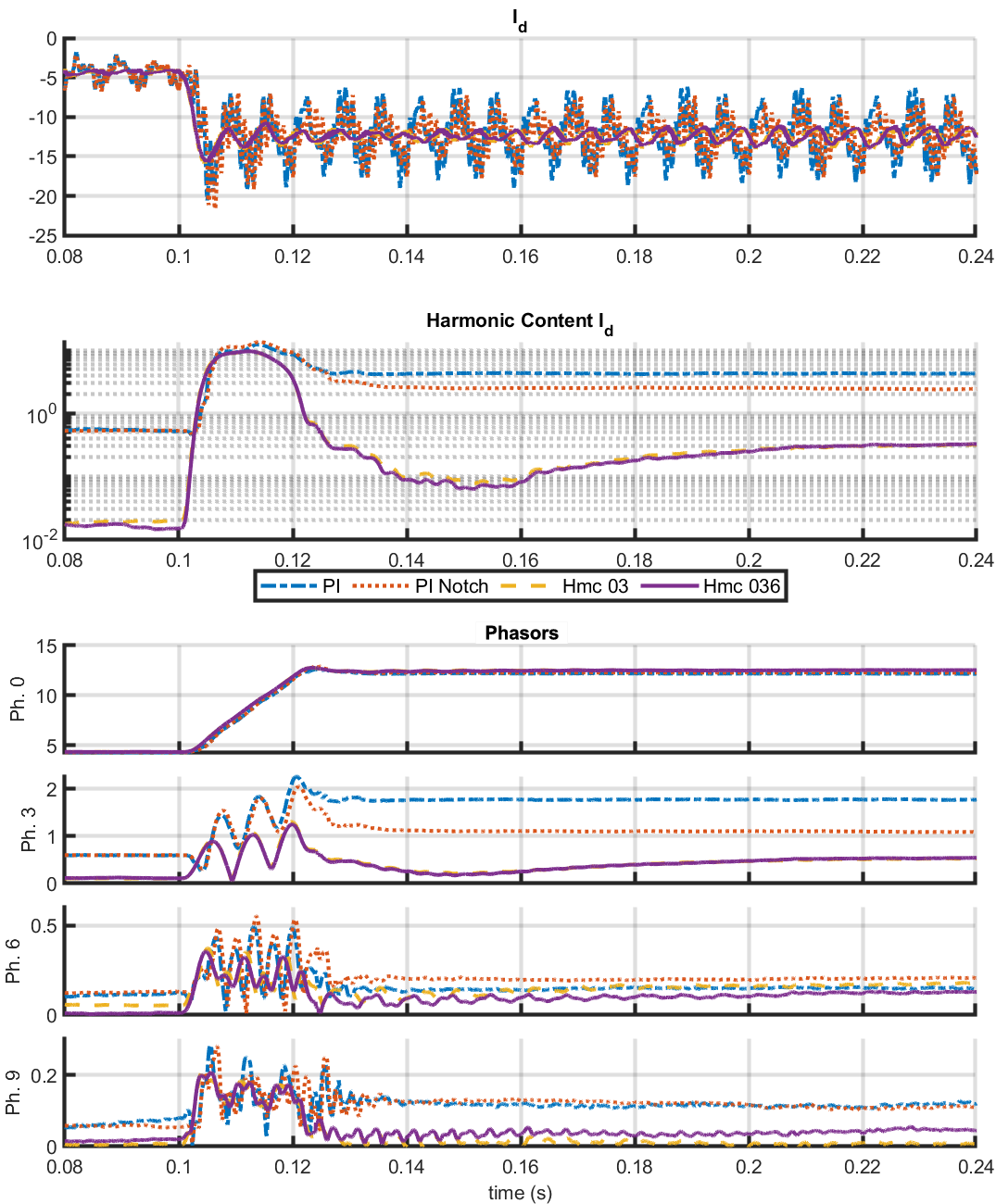}%
\label{fig:ExpAss_I_step_h3_subfig:id}}
\hfil
\subfloat[]{\includegraphics[width=1\columnwidth]{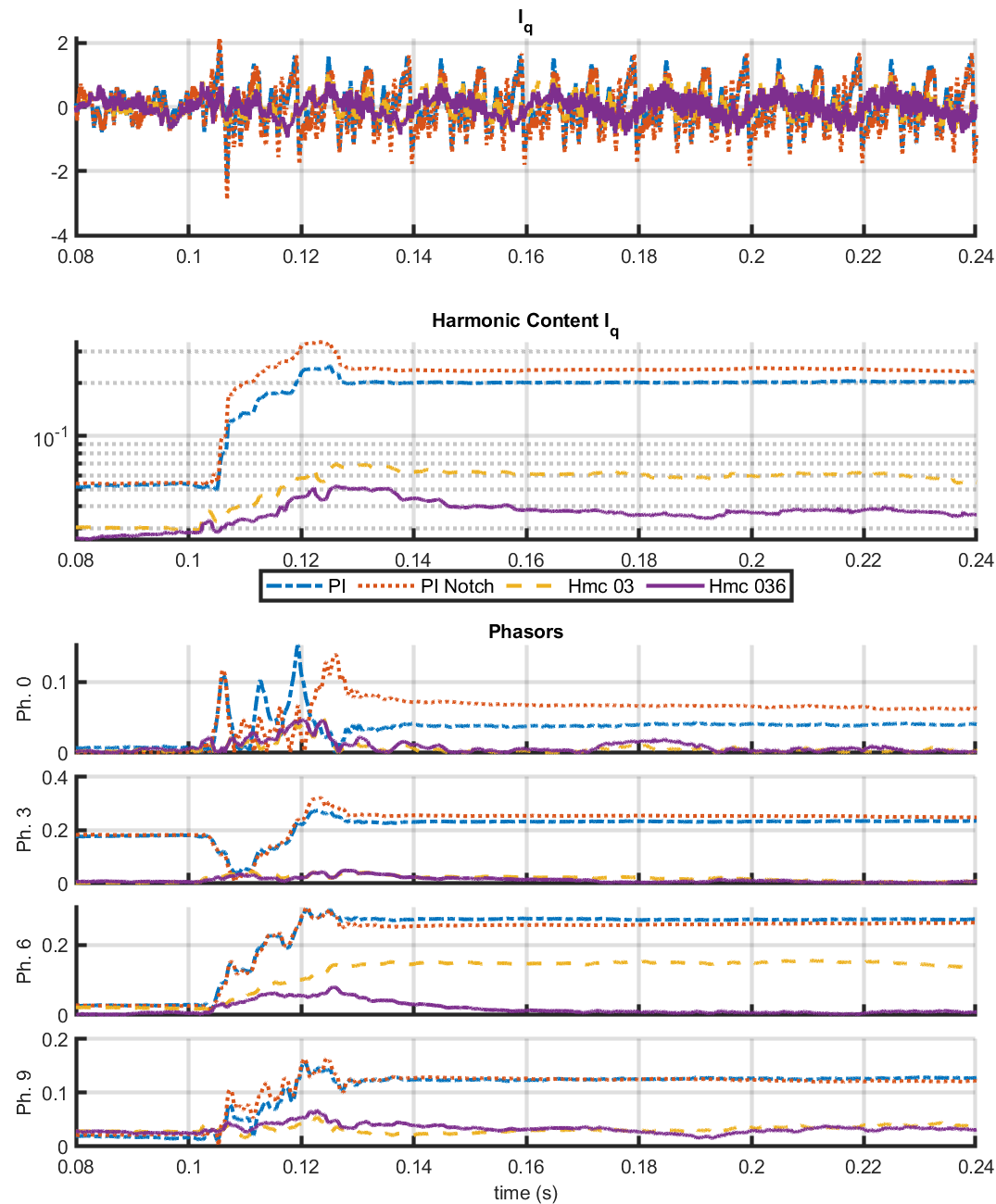}%
\label{fig:ExpAss_I_step_h3_subfig:iq}}
\vfil
\subfloat[]{\includegraphics[width=1\columnwidth]{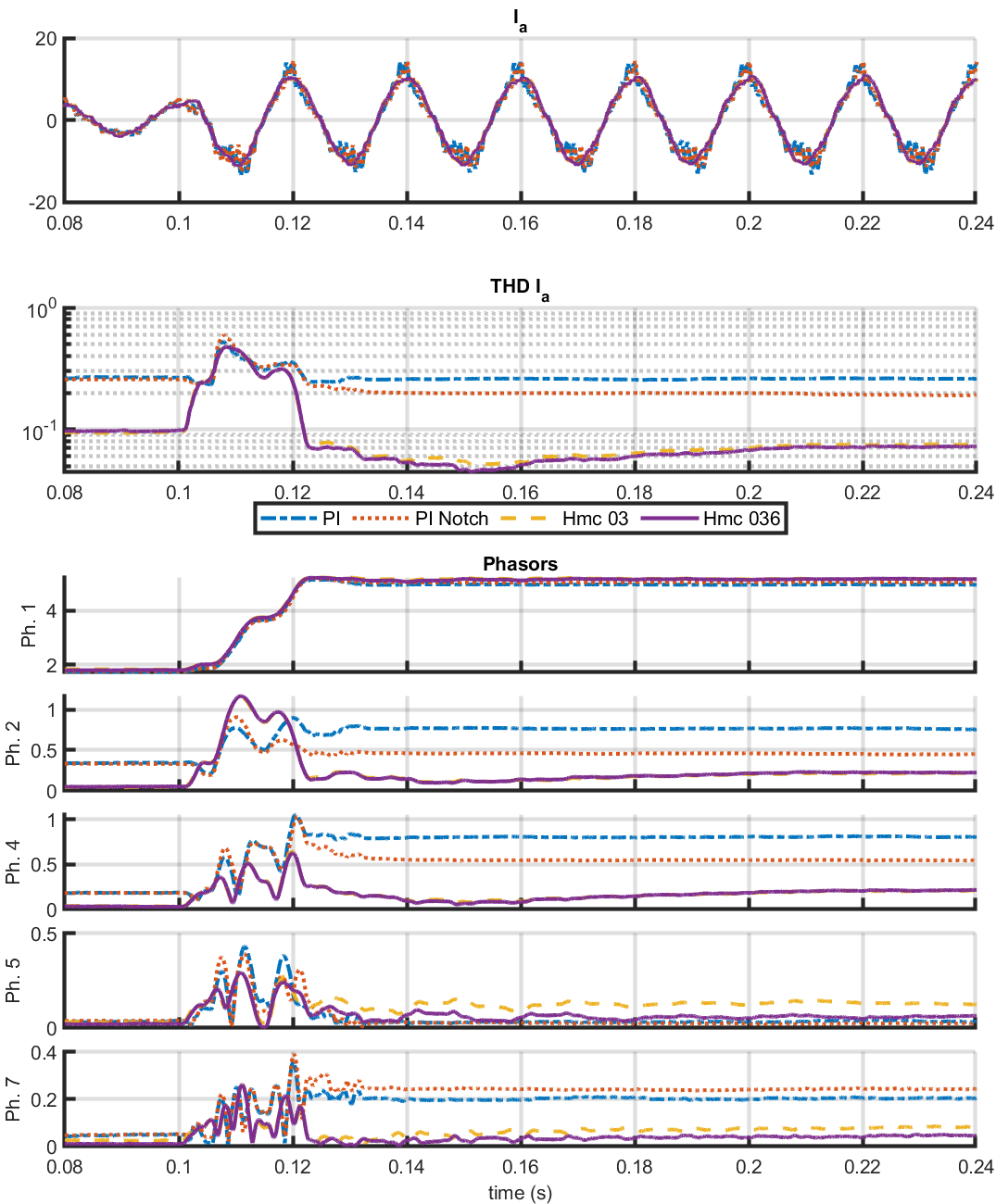}%
\label{fig:ExpAss_I_step_h3_subfig:ia}}
\hfil
\subfloat[]{\includegraphics[width=1\columnwidth]{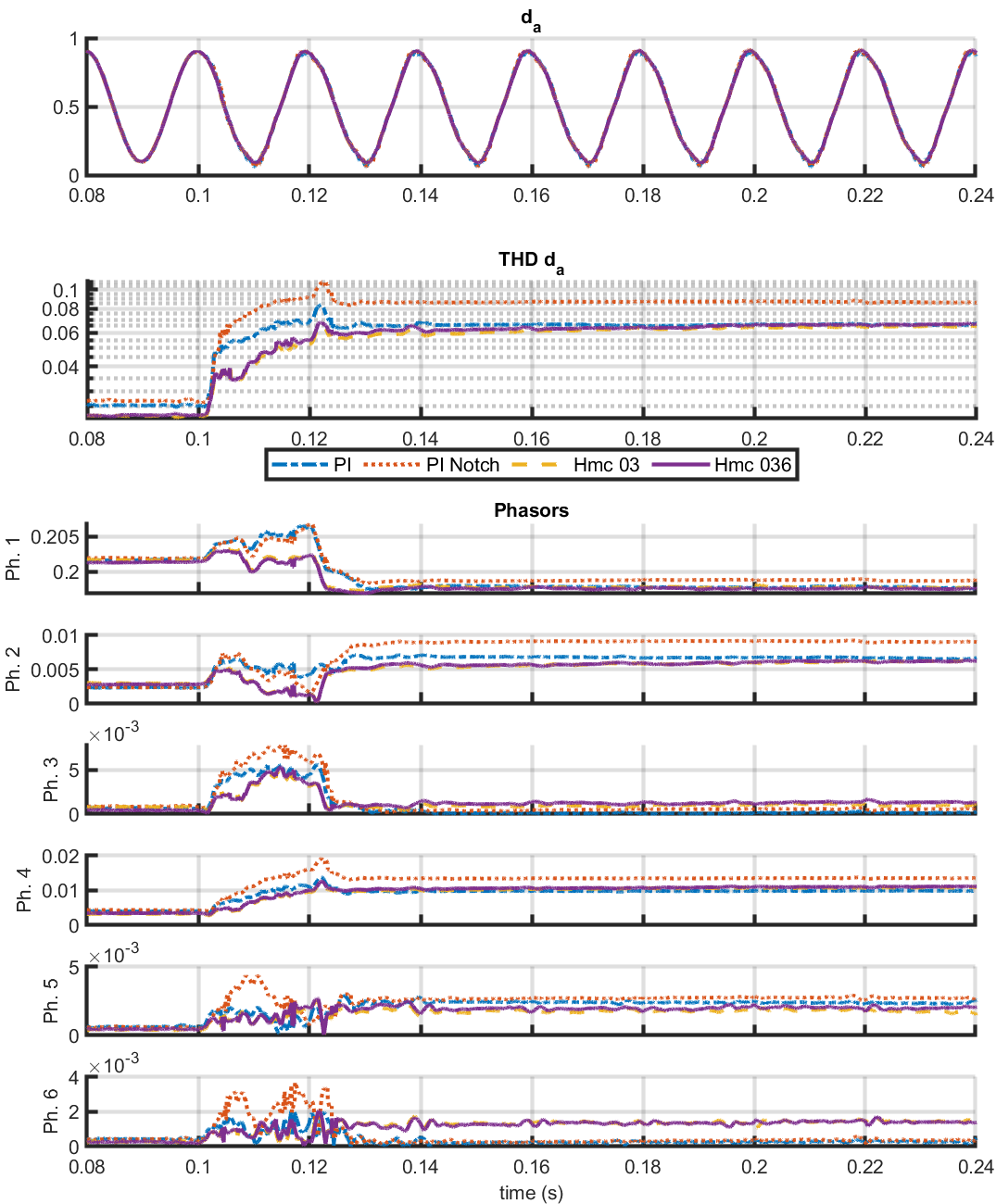}%
\label{fig:ExpAss_I_step_h3_subfig:sa}}
\caption{Load side harmonic injection: AC GRID (a) $i_d$, (b) $i_q$, (c) $i_a$ and (d) $d_a$}
\label{fig:ExpAss_I_step_h3_ACside}
\end{figure*}

 
To compare the effectiveness of our proposed harmonic control scheme with traditional PI control strategies, we consider a total of four control cases (Figures \ref{fig:ExpAss_I_step_h3_DCside} and \ref{fig:ExpAss_I_step_h3_ACside}), the following notations are used:
\begin{itemize}[nosep]
\item \textbf{PI}: The conventional PI control scheme without a notch filter, allowing observation of the system's inherent tendency to propagate harmonic content.
\item \textbf{PI Notch}: The PI control strategy augmented with a Notch filter on the $i_{dq}$ reference signals, specifically tuned to attenuate the 150 Hz component.
\item \textbf{Hmc 03}: The harmonic control scheme targeted at the 0 phasor of $v_{dc}$/$i_q$ and the 3rd phasor of $i_{dq}$, as detailed in Section \ref{whole_ctrl}.
\item \textbf{Hmc 036}: The extended harmonic control scheme, which includes the same targeting objectives as Hmc 03 and additionally aims to control the 6th phasor of $i_{dq}$.
\end{itemize}

First, let us examine the current harmonic content from the programmable current load as depicted in Figure~\ref{fig:ExpAss_I_step_h3_subfig:idcsink}. In order to ensure a meaningful comparison despite the presence of a non-ideal DC source, we took precautions to inject approximately the same $0^\text{th}$ and $3^\text{rd}$ phasors across the four experiments. It is worth noting that the non-ideal DC source introduces higher-order harmonics, particularly multiples of $150$ Hz. 

Regarding the regulation of the DC bus voltage in Figure~\ref{fig:ExpAss_I_step_h3_subfig:vdc}, both harmonic control approaches demonstrate a faster response to the disturbance. The inclusion of the $6^\text{th}$ order phasor action in the harmonic control does not have a significant impact on the dynamics. It is worth noting that, in all control strategies, small deviations from the reference (represented by the $0^\text{th}$ phasor) are observed when a harmonic disturbance is introduced. This deviation was not present in Figure~\ref{fig:ExpAss_I_Step} for the step input without harmonic injection on the load side.  

By focusing on Figures \ref{fig:ExpAss_I_step_h3_subfig:id}, \ref{fig:ExpAss_I_step_h3_subfig:iq}, and \ref{fig:ExpAss_I_step_h3_subfig:ia}, we can observe the impact of the control strategies on the transient behavior of the currents and the resulting Total Harmonic Distortion (THD). In a systematic manner, both harmonic control approaches exhibit superior performance in terms of harmonic content in the considered signals. As anticipated, the inclusion of a Notch filter provides some improvement, but its effectiveness falls short compared to the harmonic design.

Furthermore, in Figure~\ref{fig:ExpAss_I_step_h3_subfig:ia} and \ref{fig:ExpAss_I_step_h3_subfig:iq}, we notice that adding a $6th$ phasor rejection in the harmonic design effectively targets the $5th$ and $7th$ harmonic content in $i_a$ (corresponding to the $6^\text{th}$ phasor in $i_\dq$). However, in this particular case, it does not significantly contribute to further reduction of THD. It is worth noting that the mean value of $i_q$ (the $0th$ phasor) is successfully regulated to zero using the harmonic control approach, whereas both PI strategies fail to completely eliminate this component, resulting in a non-unity power factor.
Overall, the systematic comparison of the control strategies, as illustrated in these figures, confirms the superiority of the harmonic control approach in terms of harmonic content reduction and achieving a unity power factor.

It is important to note that, despite our control design, the $3^\text{rd}$ phasor of $i_d$ is not completely rejected. Indeed, our harmonic design ensures total rejection only in the case of a constant input disturbance in the harmonic domain, or equivalently, for a pure $T$-period sinusoidal signal in the time domain. However, since the actual harmonic disturbance is non-ideal and varies over time, the integral action of our control design can only guarantee attenuation of this disturbance rather than complete rejection. This limitation arises due to the dynamic and time-varying nature of the harmonic disturbance, which cannot be fully compensated for by the integral action alone.

 Finally, let us observe in Figure~\ref{fig:ExpAss_I_step_h3_subfig:sa} that the harmonic content of the control differs significantly for strategic phasor orders. Although the visual difference may be subtle, it undeniably translates into superior performance. 

\subsection{Robustness issues}
Robustness features of the proposed control strategy have been checked experimentaly. No significant degradation in experimental performance was observed when deviating the parameters ($r, L, C$) in the model \eqref{model:converter_bilmodel_abc_time} by $\pm 40\%$ from their actual experimental values and synthesizing the controller based on these deviated parameters. Although the control strategy was specifically designed for a nominal input frequency of 50Hz on the AC side and implemented with an approach that updates the $\omega$ term in the $O$ matrix (as described in \eqref{finalOLC_time}) using information from the PLL (as shown in \eqref{pll}), stability has been preserved within the operating range of 30Hz to 80Hz. 

\section*{Conclusion}
This paper presents the construction of a bilinear infinite harmonic model for an AC/DC converter connected to a DC load. Leveraging this model, a globally stabilizing state feedback controller is designed in the harmonic domain. Importantly, we derive the corresponding time domain counterpart of the controller, ensuring the same guaranteed stability property, even in the presence of control saturation. Furthermore, the effectiveness of integral actions in tracking periodic trajectories and mitigating the impact of harmonic disturbances is demonstrated. The proposed control strategy has been implemented in both simulation and on an experimental test bench. 
The results showcase the real-time operation of the control approach at a conventional sampling frequency, validating its feasibility for practical implementation. Moreover, the obtained experimental performance highlights the favorable outcomes achieved in terms of total harmonic distortion and harmonic content values. These positive results are observed under both nominal operating conditions and in the presence of harmonic disturbances on the load side. This confirms the potential and promise of the harmonic-based control framework in effectively managing and controlling the harmonic content within energy management applications.


\bibliography{IEEEabrv,bibtex/bib/better_bib_v1}

\begin{thebibliography}{10}
\providecommand{\url}[1]{#1}
\csname url@samestyle\endcsname
\providecommand{\newblock}{\relax}
\providecommand{\bibinfo}[2]{#2}
\providecommand{\BIBentrySTDinterwordspacing}{\spaceskip=0pt\relax}
\providecommand{\BIBentryALTinterwordstretchfactor}{4}
\providecommand{\BIBentryALTinterwordspacing}{\spaceskip=\fontdimen2\font plus
\BIBentryALTinterwordstretchfactor\fontdimen3\font minus
  \fontdimen4\font\relax}
\providecommand{\BIBforeignlanguage}[2]{{%
\expandafter\ifx\csname l@#1\endcsname\relax
\typeout{** WARNING: IEEEtran.bst: No hyphenation pattern has been}%
\typeout{** loaded for the language `#1'. Using the pattern for}%
\typeout{** the default language instead.}%
\else
\language=\csname l@#1\endcsname
\fi
#2}}
\providecommand{\BIBdecl}{\relax}
\BIBdecl

\bibitem{mollerstedtOutControlBecause2000}
E.~Mollerstedt and B.~Bernhardsson, ``Out of {{Control Because}} of
  {{Harmonics}}: {{An Analysis}} of the {{Harmonic Response}} of an {{Inverter
  Locomotive}},'' \emph{IEEE Control Systems}, vol.~20, no.~4, pp. 70--81,
  2000.

\bibitem{liserreMultipleHarmonicsControl2006}
M.~Liserre, R.~Teodorescu, and F.~Blaabjerg, ``Multiple harmonics control for
  three-phase grid converter systems with the use of {{PI-RES}} current
  controller in a rotating frame,'' \emph{IEEE Transactions on Power
  Electronics}, vol.~21, no.~3, pp. 836--841, May 2006.

\bibitem{teodorescuProportionalresonantControllersFilters2006}
R.~Teodorescu, F.~Blaabjerg, M.~Liserre, and P.~C. Loh, ``Proportional-resonant
  controllers and filters for grid-connected voltage-source converters,''
  \emph{IEE Proceedings: Electric Power Applications}, vol. 153, no.~5, pp.
  750--762, 2006.

\bibitem{yuanStationaryframeGeneralizedIntegrators2002}
X.~Yuan, W.~Merk, H.~Stemmler, and J.~Allmeling, ``Stationary-frame generalized
  integrators for current control of active power filters with zero
  steady-state error for current harmonics of concern under unbalanced and
  distorted operating conditions,'' \emph{IEEE Transactions on Industry
  Applications}, vol.~38, no.~2, pp. 523--532, Mar. 2002.

\bibitem{almerDynamicPhasorModel2015}
S.~Alm{\'e}r, S.~Mari{\'e}thoz, and M.~Morari, ``Dynamic {{Phasor Model
  Predictive Control}} of {{Switched Mode Power Converters}},'' \emph{IEEE
  Transactions on Control Systems Technology}, vol.~23, no.~1, pp. 349--356,
  Jan. 2015.

\bibitem{chandrasekarAdaptiveHarmonicSteadyState2006}
J.~Chandrasekar, L.~Liu, D.~Patt, P.~Friedmann, and D.~Bernstein, ``Adaptive
  {{Harmonic Steady-State Control}} for {{Disturbance Rejection}},'' \emph{IEEE
  Transactions on Control Systems Technology}, vol.~14, no.~6, pp. 993--1007,
  Nov. 2006.

\bibitem{kamaldarAdaptiveHarmonicControl2020}
M.~Kamaldar and J.~B. Hoagg, ``Adaptive {{Harmonic Control}} for {{Rejection}}
  of {{Sinusoidal Disturbances Acting}} on an {{Unknown System}},'' \emph{IEEE
  Transactions on Control Systems Technology}, vol.~28, no.~2, pp. 277--290,
  Mar. 2020.

\bibitem{ricoDynamicHarmonicEvolution2003}
J.~J. Rico, M.~Madrigal, and E.~Acha, ``Dynamic harmonic evolution using the
  extended harmonic domain,'' \emph{IEEE Transactions on Power Delivery},
  vol.~18, no.~2, pp. 587--594, Apr. 2003.

\bibitem{karamiSinglePhaseModelingApproach2017}
E.~Karami, M.~Madrigal, G.~B. Gharehpetian, K.~Rouzbehi, and P.~Rodriguez,
  ``Single-{{Phase Modeling Approach}} in {{Dynamic Harmonic Domain}},''
  \emph{IEEE Transactions on Power Systems}, vol.~33, no.~1, pp. 257--267, Mar.
  2017.

\bibitem{kotianDynamicPhasorModelling2016}
S.~M. Kotian and K.~N. Shubhanga, ``Dynamic phasor modelling and simulation,''
  \emph{12th IEEE International Conference Electronics, Energy, Environment,
  Communication, Computer, Control: (E3-C3), INDICON 2015}, Mar. 2016.

\bibitem{javaidArbitraryOrderGeneralized2015}
U.~Javaid and D.~Duji{\'c}, ``Arbitrary order generalized state space average
  modeling of switching converters,'' in \emph{2015 {{IEEE Energy Conversion
  Congress}} and {{Exposition}} ({{ECCE}})}, Sep. 2015, pp. 6399--6406.

\bibitem{sandersGeneralizedAveragingMethod1991}
S.~R. Sanders, J.~M. Noworolski, X.~Z. Liu, and G.~C. Verghese, ``Generalized
  {{Averaging Method}} for {{Power Conversion Circuits}},'' \emph{IEEE
  Transactions on Power Electronics}, vol.~6, no.~2, pp. 251--259, 1991.

\bibitem{qinGeneralizedAverageModeling2012}
H.~Qin and J.~W. Kimball, ``Generalized {{Average Modeling}} of {{Dual Active
  Bridge DC}}\textendash{{DC Converter}},'' \emph{IEEE Transactions on Power
  Electronics}, vol.~27, no.~4, pp. 2078--2084, Apr. 2012.

\bibitem{kwonHarmonicInteractionAnalysis2015}
J.~Kwon, X.~Wang, C.~L. Bak, and F.~Blaabjerg, ``Harmonic interaction analysis
  in grid connected converter using harmonic state space ({{HSS}}) modeling,''
  \emph{Conference Proceedings - IEEE Applied Power Electronics Conference and
  Exposition - APEC}, vol. 2015-May, no. May, pp. 1779--1786, May 2015.

\bibitem{kwonFrequencyDomainModelingSimulation2017}
J.~Kwon, X.~Wang, F.~Blaabjerg, and C.~L. Bak, ``Frequency-{{Domain Modeling}}
  and {{Simulation}} of {{DC Power Electronic Systems Using Harmonic State
  Space Method}},'' \emph{IEEE Transactions on Power Electronics}, vol.~32,
  no.~2, pp. 1044--1055, Feb. 2017.

\bibitem{ormrodHarmonicStateSpace2013}
J.~E. Ormrod, ``Harmonic state space modelling of voltage source converters,''
  Ph.D. dissertation, University of Canterbury, {Christchurch, New Zealand},
  2013.

\bibitem{blinNecessarySufficientConditions2022}
N.~Blin, P.~Riedinger, J.~Daafouz, L.~Grimaud, and P.~Feyel, ``Necessary and
  {{Sufficient Conditions}} for {{Harmonic Control}} in {{Continuous Time}},''
  \emph{IEEE Transactions on Automatic Control}, vol.~67, no.~8, pp.
  4013--4028, Aug. 2022.

\bibitem{gaviriaRobustControllerFullbridge2005}
C.~Gaviria, E.~Fossas, and R.~Grino, ``Robust controller for a full-bridge
  rectifier using the {{IDA}} approach and {{GSSA}} modeling,'' \emph{IEEE
  Transactions on Circuits and Systems I: Regular Papers}, vol.~52, no.~3, pp.
  609--616, Mar. 2005.

\bibitem{ghitaHarmonicStateSpace2017}
I.~Ghita, P.~Kvieska, D.~Beauvois, and E.~Godoy, ``Harmonic state space
  feedback control for {{AC}}/{{DC}} power converters,'' \emph{Proceedings of
  the American Control Conference}, pp. 4087--4092, Jun. 2017.

\bibitem{riedingerSolvingInfiniteDimensionalHarmonic2022}
P.~Riedinger and J.~Daafouz, ``Solving {{Infinite-Dimensional Harmonic
  Lyapunov}} and {{Riccati Equations}},'' \emph{IEEE Transactions on Automatic
  Control}, 2022.

\bibitem{riedingerHarmonicPolePlacement2022}
------, ``Harmonic {{Pole Placement}},'' in \emph{2022 {{IEEE}} 61st
  {{Conference}} on {{Decision}} and {{Control}} ({{CDC}})}, Dec. 2022, pp.
  5505--5510.

\bibitem{vernereySolvingInfinitedimensionalToeplitz2023}
F.~Vernerey, P.~Riedinger, and J.~Daafouz, ``On solving infinite-dimensional
  {{Toeplitz Block LMIs}},'' May 2023, [Online]
  \href{http://arxiv.org/abs/2303.08465}{arXiv:2303.08465}.

\bibitem{zhouModelingControlThreePhase2018}
D.~Zhou, Y.~Song, and F.~Blaabjerg, ``Modeling and {{Control}} of {{Three-Phase
  AC}}/{{DC Converter Including Phase-Locked Loop}},'' in \emph{Control of
  {{Power Electronic Converters}} and {{Systems}}}.\hskip 1em plus 0.5em minus
  0.4em\relax {Academic Press}, Jan. 2018, pp. 117--151.

\bibitem{mazencAddingIntegrationsSaturated1996}
F.~Mazenc and L.~Praly, ``Adding integrations, saturated controls, and
  stabilization for feedforward systems,'' \emph{IEEE Transactions on Automatic
  Control}, vol.~41, no.~11, pp. 1559--1578, 1996.

\bibitem{simonRobustRegulationPower2023}
T.~Simon, M.~Giaccagli, J.-F. Tregouet, D.~Astolfi, V.~Andrieu, H.~Morel, and
  X.~{Lin-Shi}, ``Robust {{Regulation}} of a {{Power Flow Controller}} via
  {{Nonlinear Integral Action}},'' \emph{IEEE Transactions on Control Systems
  Technology}, pp. 1--13, Jan. 2023.

\bibitem{wereleyLinearTimePeriodic1991}
N.~M. Wereley and S.~R. Hall, ``Linear time periodic systems: {{Transfer}}
  function, poles, transmission zeroes and directional properties,''
  \emph{Proceedings of the American Control Conference}, vol.~2, pp.
  1179--1184, 1991.

\bibitem{zhouSpectralCharacteristicsEigenvalues2004}
J.~Zhou, T.~Hagiwara, and M.~Araki, ``Spectral characteristics and eigenvalues
  computation of the harmonic state operators in continuous-time periodic
  systems,'' \emph{Systems \& Control Letters}, vol.~53, no.~2, pp. 141--155,
  Oct. 2004.

\bibitem{sritaModelingSimulationDevelopment2022}
S.~Srita, S.~Somkun, T.~Kaewchum, W.~Rakwichian, P.~Zacharias, U.~Kamnarn,
  J.~Thongpron, D.~Amorndechaphon, and M.~Phattanasak, ``Modeling,
  {{Simulation}} and {{Development}} of {{Grid-Connected Voltage Source
  Converter}} with {{Selective Harmonic Mitigation}}: {{HiL}} and
  {{Experimental Validations}},'' \emph{Energies}, vol.~15, no.~7, p. 2535,
  Jan. 2022.

\bibitem{zhangMethodBasedGeneralized1999}
B.~Zhang, ``The method based on a generalized dq/sub k/ coordinate transform
  for current detection of an active power filter and power system,'' in
  \emph{30th {{Annual IEEE Power Electronics Specialists Conference}}.
  {{Record}}. ({{Cat}}. {{No}}.{{99CH36321}})}, vol.~1, Jul. 1999, pp. 242--248
  vol.1.

\bibitem{beneuxAdaptiveStabilizationSwitched2019}
G.~Beneux, P.~Riedinger, J.~Daafouz, and L.~Grimaud, ``Adaptive stabilization
  of switched affine systems with unknown equilibrium points: {{Application}}
  to power converters,'' \emph{Automatica}, vol.~99, pp. 82--91, Jan. 2019.

\end{thebibliography}
\bibliographystyle{IEEEtran}

\begin{IEEEbiography}[{\includegraphics[width=1in,height=1.25in,clip,keepaspectratio]{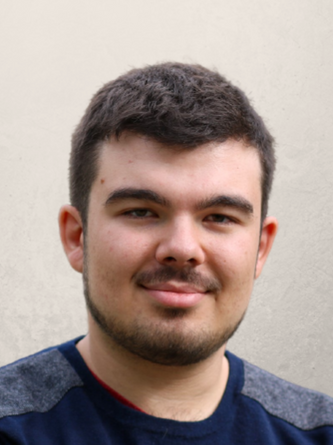}}]{Maxime Grosso} received an Engineering Diploma and a Master's degree in Automatic Control from Ecole Centrale de Lyon (France) in 2021. Since 2022, he has been a PhD Student at CRAN-CNRS Université de Lorraine (Nancy, France) and is affiliated with Safran Electronics and Defense (Massy, France). His research interests include energy conversion, power system modeling and control, and harmonic modeling and control.
\end{IEEEbiography}\vskip -2\baselineskip plus -1fil
\begin{IEEEbiography}[{\includegraphics[width=1in,height=1.25in,clip,keepaspectratio]{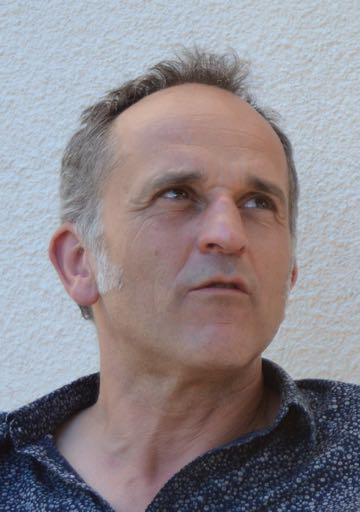}}]{Pierre Riedinger}
is a Full Professor at University de Lorraine (France) and researcher at CRAN CNRS. He received his M.Sc. degree in Applied Mathematics from the University Joseph
Fourier, Grenoble in 1993 and the Ph.D. degree in Automatic Control in 1999 from the Institut
National Polytechnique de Lorraine (INPL). He got the French Habilitation degree from the INPL
in 2010. His current research interests include control theory and optimization of systems with their applications in electrical and power systems.
\end{IEEEbiography}\vskip -2\baselineskip plus -1fil
\begin{IEEEbiography}[{\includegraphics[width=1in,height=1.25in,clip,keepaspectratio]{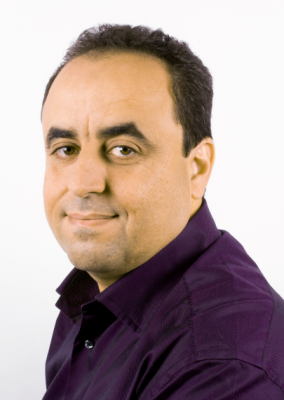}}]{Jamal Daafouz}
is a Full Professor at University de Lorraine (France) and researcher at CRAN CNRS. In 1994, he received a Ph.D. in Control from INSA Toulouse, in 1997. He also received the "Habilitation à Diriger des Recherches" from INPL (Université de Lorraine), Nancy, in 2005. His research interests include analysis, observation and control of switched systems and networked systems with a particular interest for convex based optimisation methods. In 2010, he was appointed as a junior member of the Institut Universitaire de France (IUF). He served as an associate editor of the journals: Automatica, IEEE Trans. on Automatic Control, European Journal of Control and Non linear Analysis and Hybrid Systems. He is senior editor of IEEE Control Systems Letters.
\end{IEEEbiography}\vskip -2\baselineskip plus -1fil
\begin{IEEEbiography}[{\includegraphics[width=1in,height=1.25in,clip]{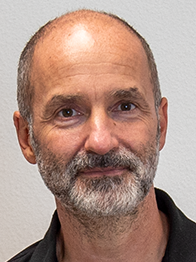}}]{Serge Pierfederici}
received the Dipl.-Ing. degree in electrical engineering from the Ecole Nationale Supérieure d’Electricité et de Mécanique (ENSEM) of Institut National Polytechnique de Lorraine (INPL), Nancy, France, in 1994, and the Ph.D. degree from the Institut National Polytechnique de Lorraine (INPL), Nancy, in 1998. Since 1999, he's working at the Lorraine University, where he is currently a Full Professor. His research activities deals with the stability study of distributed power system, the control of DC and AC microgrids and the design of power electronic converters for specific applications like fuel cells systems, electrolyser.
\end{IEEEbiography}\vskip -2\baselineskip plus -1fil
\begin{IEEEbiography}[{\includegraphics[width=1in,height=1in,clip]{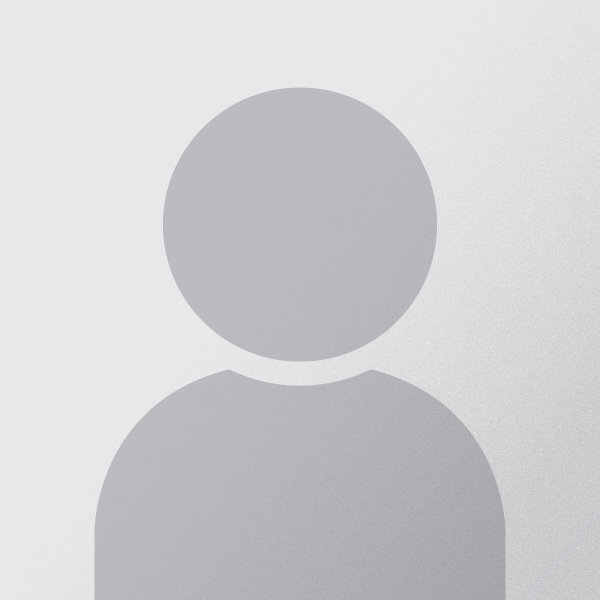}}]
{Blaise Lapôtre} (photograph not available at the time of publication)
is a R\&T engineer in Safran Electronics and Defense (SED). Expert in electric motor design, he graduated from "École nationale supérieure d’électricité et de mécanique" (ENSEM) in 2012. He received a PhD in 2016 from IAEM (Université de Lorraine) in Nancy. 
\end{IEEEbiography}\vskip -2\baselineskip plus -1fil
\begin{IEEEbiography}[{\includegraphics[width=1in,height=1.25in,clip,keepaspectratio]{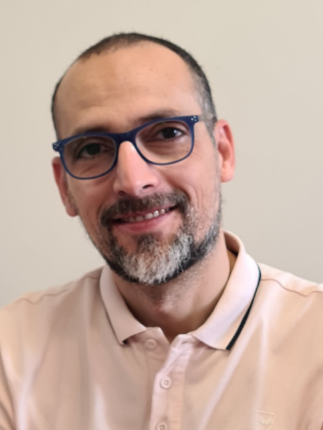}}]{Hicham Janati-Idrissi}  is a control expert in Safran Electronics and Defence (SED) in electromechanical systems. He graduated from USTL Lille in 1999 and joined Safran Group to work on SaM146 engine control system. He has also worked on several themes in the context of industrial projects such as the prognostic and health monitoring of aeronautical systems and the modeling and control of fuel cell systems. He is currently working on flight control systems based on impedance control theory for fly by wire equipment integrated into aircraft cockpits.
\end{IEEEbiography}

\end{document}